\documentclass[10pt,reqno,final]{article}

\usepackage{amsmath,amsfonts,amssymb,amsthm,version}
\usepackage{mathrsfs,fancybox,pifont}
\usepackage{graphicx}
\usepackage[nocompress]{cite}
\usepackage{url,hyperref}
\usepackage[notcite,notref]{showkeys}
\usepackage{color}
\usepackage{subfigure,multirow}
\usepackage{epstopdf}
\usepackage{cases}
\usepackage[shortlabels]{enumitem}
\usepackage{mathtools}
\usepackage{algorithm,algorithmic}
\usepackage{authblk}
\usepackage{lipsum}
\usepackage{tikz-cd}
\usepackage{braket}

\usepackage{import}
\usepackage{xifthen}
\usepackage{pdfpages}
\usepackage{transparent}
\usepackage{longtable}
\usepackage{booktabs}

\allowdisplaybreaks

\renewcommand {\thefootnote}{\fnsymbol{footnote}}

\newcommand{\dd}{\mathop{}\!\mathrm{d}}

\numberwithin{equation}{section}
\numberwithin{figure}{section}
\numberwithin{table}{section}

\theoremstyle{plain}
\newtheorem{theorem}{Theorem}[section]
\newtheorem{lemma}[theorem]{Lemma}
\newtheorem{corollary}[theorem]{Corollary}
\newtheorem{proposition}[theorem]{Proposition}

\theoremstyle{definition}
\newtheorem{definition}[theorem]{Definition}

\theoremstyle{remark}
\newtheorem{remark}[theorem]{Remark}

\title{Concavity of Normalized Generalized Brézin–Gross–Witten Correlators and Its Applications}
\author[1]{Chenglang Yang\thanks{yangcl@whu.edu.cn}}
\author[2]{Xuhui Zhang\thanks{zhangxh.math@gmail.com}}

\affil[1]{Institute for Math and AI, Wuhan University, Wuhan, China}
\affil[2]{School of Mathematics, Sun Yat-sen University, Guangzhou, China}

\date{}

\begin{document}
\maketitle
\begin{abstract}
  In this paper,
  we study the generalized Br\'ezin--Gross--Witten model using its BKP integrability.
  By virtue of the formula for connected \(n\)-point functions of tau-functions of the BKP hierarchy obtained by Wang and the first-named author,
  we prove that the normalized generalized BGW correlators are strictly concave for fixed genus and fixed total degree.
  As applications, we establish the strict Schur-concavity and derive upper and lower bounds for these correlators.
  We also prove the strict monotonicity of normalized ordinary BGW correlators,
  which resolves a conjecture by Guo--Norbury--Yang--Zagier.
\end{abstract}

{\let\thefootnote\relax}

\section{Introduction}
\subsection{Background}
The Brézin–Gross–Witten (BGW) model was introduced as a matrix model by Br\'ezin--Gross \cite{BG80} and Gross--Witten \cite{GW80} to study lattice gauge theory.
Its partition function is given by
\begin{equation}
  Z_{BGW}(A;\hbar)=\int_{U(M)} \exp \left(\frac{1}{\hbar}\operatorname{Tr}(A^{\dagger}U+AU^{\dagger})\right)\dd U,
\end{equation}
where \(U(M)\) is the group of \(M\times M\) unitary matrices, \(A\) is an arbitrary complex \(M\times M\) source matrix, and \(\dd U\) denotes the normalized Haar measure on \(U(M)\).
When we use the Miwa variables $t_k=\text{Tr}((A^\dagger A)^{-k/2})/k$ and consider $M\rightarrow\infty$,
the partition function \(Z_{BGW}\) becomes a tau function of the KdV hierarchy and satisfies the Virasoro constraints (see \cite{MMS96,Alex18,GN92} etc.). Recently, the BKP integrability and an elegant Schur Q-polynomial expansion of this model have been established \cite{Alex21a,Alex21b,Alex23,LY22,LY25},
which suggests applying the BKP method to study the properties of the BGW model.

The BGW correlators are coefficients of the expansion of the partition function of the BGW model.
From the geometric viewpoint, the BGW correlators admit an intersection-theoretic interpretation on the moduli space of stable curves, analogous to the Witten–Kontsevich case \cite{Witten90,Kont92}.
Norbury \cite{Nor23} introduced classes \(\Theta_{g,n}\in H^{4g-4+2n}(\overline{\mathcal{M}}_{g,n},\mathbb{Q})\) and conjectured that the BGW correlators are given by
\begin{equation}\label{eqn:BGW<> def}
  \langle \tau_{d_{1}}\cdots \tau_{d_{n}} \rangle_{g}^{\Theta}= \int_{\overline{\mathcal{M}}_{g,n}}\Theta_{g,n}\psi_{1}^{d_{1}}\cdots \psi_{n}^{d_{n}},
\end{equation}
where $\overline{\mathcal{M}}_{g,n}$ is the moduli space of stable genus $g$ curves with $n$ distinct smooth marked points,
and $\psi_i\in H^{2}(\overline{\mathcal{M}}_{g,n},\mathbb{Q})$ is the first Chern class of the $i$-th cotangent line bundle.
This conjecture was later proved by Chidambaram–Garcia-Failde–Giacchetto \cite{CGG25}.
In recent years,
many advances have been achieved on the topic of large-genus asymptotics of intersection numbers of the moduli spaces of curves (see \cite{Mir13,MZ15,DGZZ21,Agg21,GY24,GYZ26,EGGGL26} and references therein).
In particular,
Guo--Norbury--Yang--Zagier \cite{GNYZ25} investigated several properties of suitably normalized BGW correlators and made significant progress on the uniform large-genus asymptotics of these correlators.
They also proposed some conjectures regarding the upper bound and monotonicity of the normalized BGW correlators.

The BGW model admits a one-parameter deformation using the matrix model \cite{Penner88,MMS96}.
The corresponding partition function is called the generalized Br\'ezin--Gross--Witten model and was introduced by Mironov--Morozov--Semenoff \cite{MMS96}.
The generalized BGW model has also been studied from both integrable and enumerative-geometric perspectives (see \cite{Alex23,Alex21b,BR19,DYZ21,LY22,YZ21,YZ23,Wang22,XY25,WYZ25} and references therein).
On the type B integrable system side, 
it was proved that the  partition function of the generalized BGW model is a tau-function of the BKP hierarchy after rescaling, and admits an explicit expansion in terms of Schur \(Q\)-functions \cite{LY22,Alex23,Alex21b,Alex21a}.
Recently,
Wang--Yang \cite{WY22} and Zhang--Zhou \cite{ZZ26} derived two types of formulae to compute the connected \(n\)-point functions of tau-functions of the BKP hierarchy and used them to compute the correlators of the generalized BGW model.
 Building on these results, Wang--Yang--Zhang \cite{WYZ25} derived the corresponding spectral curve and quantum spectral curve of type B using the emergent geometry.
 On the enumerative-geometric side,
 Yang--Zhang \cite{YZ24} established the Hodge--BGW correspondence and obtained an ELSV-type formula expressing generalized BGW correlators in terms of cubic Hodge integrals. Alexandrov--Norbury \cite{AN24} introduced the spin class \(\Omega_{g,n+2h}^{(1^{n},0^{2h})}\in H^{4g-4+2n+2h}(\overline{\mathcal{M}}_{g,n+2h},\mathbb{Q})\) and showed that the generalized BGW correlators admit the intersection-theoretic expression:
\begin{equation}\label{eq:intro-gBGW-correlator}
  \langle \tau_{d_{1}}\cdots \tau_{d_{n}} \rangle_{g}^{\Omega,h}=\frac{1}{(2h)!}\int_{\overline{\mathcal{M}}_{g,n+2h}}\Omega_{g,n+2h}^{(1^{n},0^{2h})}\prod_{i=1}^n\psi_{i}^{d_i}.
\end{equation}
For \(h=0\), this construction recovers the ordinary BGW intersection numbers \(\langle \tau_{d_{1}}\cdots \tau_{d_{n}} \rangle_{g}^{\Theta}\).
In this paper, we will adopt the notation \eqref{eqn:BGW<> def}, \eqref{eq:intro-gBGW-correlator} and use the connected $n$-point function of tau-functions of the BKP hierarchy to study the strict concavity of the normalized generalized BGW correlators and the strict monotonicity of the BGW correlators.

\subsection{Main results and organization}
We define the normalized generalized BGW correlators \(C_{h}\) by
\begin{equation}\label{eq:intro-C-h}
  C_{h}(d_{1},\dots,d_{n})
  =\frac{2^{2\sum_{i=1}^{n}d_{i}+1}\prod_{i=1}^{n}(2d_{i}+1)!!}{\left(2\sum_{i=1}^{n}d_{i}+n-1\right)!}\langle \tau_{d_{1}}\cdots \tau_{d_{n}} \rangle_{g}^{\Omega,h},
\end{equation}
where the genus $g$ is determined by the relation \(h=\sum_{i=1}^{n}d_{i}-g+1\).
The first main result is the following strict concavity of the normalized generalized BGW correlators.
\begin{theorem}\label{thm:intro-concavity}
  Let \(s \in \mathbb{N}_{+}\) and \(h\in \{ 0,\dots,s+1 \}\). The normalized generalized BGW correlators \(C_{h}\) defined in \eqref{eq:intro-C-h} on the set of partitions of size \(s\) are strictly concave.
  That means, for every partition \(\mathbf{d}=(d_{1},\dots,d_{n})\) of size \(s\) and distinct indices \(i,j\in \{ 1,\dots,n \}\) satisfying \(d_{i},d_{j}>0\), 
  we have
  \begin{equation}\label{eq:intro-concave-Ch}
    C_{h}(\mathbf{d}+e_{i}-e_{j})+C_{h}(\mathbf{d}+e_{j}-e_{i})<2 C_{h}(\mathbf{d}),
  \end{equation}
  where \(e_{i}=(\delta_{i,j})_{j=1,\dots,n}\) is the \(i\)-th standard basis vector.
\end{theorem}

As the first application of Theorem~\ref{thm:intro-concavity},
we obtain the strict Schur-concavity of the normalized generalized BGW correlators and, as an immediate consequence, derive the following bounds for them.
\begin{theorem}\label{thm:intro-bound}
  Let \(s \in \mathbb{N}_{+}\) and \(0\leq h\leq s+1\). The normalized generalized BGW correlators \(C_{h}\) defined on the set of partitions of size \(s\) with positive entries are strictly Schur-concave. Consequently, every partition \(\mathbf{d}\) of size \(s\) with \(d_{i}\in \mathbb{N}_{+}\) satisfies
  \begin{equation}\label{eqn:bound for gBGW}
    C_{h}(s)\leq C_{h}(\mathbf{d})\leq C_{h}(1^{s}).
  \end{equation}
  See \S\ref{sec:schur-concavity} for more details.
\end{theorem}

Note that when \(h=0\), these bounds \eqref{eqn:bound for gBGW} recover the known lower bound and prove the conjectured upper bound for normalized BGW correlators \cite{GNYZ25}.

Moreover,
in the proof of Theorem \ref{thm:intro-concavity},
we will provide explicit expressions for the second difference of the normalized generalized BGW correlators \(C_{h}(\mathbf{d}+e_{i}-e_{j})+C_{h}(\mathbf{d}+e_{j}-e_{i})-2C_{h}(\mathbf{d})\),
which enable us to prove the strict monotonicity conjecture of Guo--Norbury--Yang--Zagier \cite{GNYZ25}.
To state this result, we recall the colexicographic order of partitions.
For two partitions \(\mathbf{d}=(d_{1}\leq\cdots\leq d_{l})\) and \(\mathbf{d}'=(d_{1}'\leq\cdots\leq d_{l'}')\) of equal size with all entries positive, we write \(\mathbf{d}\prec \mathbf{d}'\) if either \(l<l'\), or \(l=l'\) and \(d_{j}<d_{j}'\) for the first \(j\) such that \(d_{j}\neq d_{j}'\).
Then we have the following theorem on the monotonicity of the normalized BGW correlators.
\begin{theorem}[Conjecture~2 in \cite{GNYZ25}] \label{thm:intro-monotonicity}
  For a fixed size \(s \in \mathbb{N}_{+}\), the normalized BGW correlators \(C_{0}\) are strictly increasing with respect to the colexicographic order.
  That is to say,
  for any two partitions \(\mathbf{d}, \mathbf{d}'\) of equal size \(s\) with all entries positive,
  if \(\mathbf{d}\prec \mathbf{d}'\) under the colexicographic order,
  we have
  \begin{equation}
  	C_{0}(\mathbf{d})<C_{0}(\mathbf{d}').
  \end{equation}
\end{theorem}

The rest of this paper is organized as follows. In \S\ref{sec:preliminaries}, we first recall some preliminaries on partitions and on the generalized BGW model.
In \S\ref{sec:concavity}, we prove the strict concavity of the normalized generalized BGW correlators,
which corresponds to Theorem \ref{thm:intro-concavity}. 
As applications,
in \S\ref{sec:schur-concavity} we prove Theorem \ref{thm:intro-bound},
which gives the strict Schur-concavity and bounds for the normalized generalized BGW correlators.
In \S\ref{sec:strict-monotonicity}, we prove Theorem \ref{thm:intro-monotonicity},
which gives the strict monotonicity of the normalized BGW correlators.

\section{Preliminaries}\label{sec:preliminaries}
\subsection{Integer partitions}\label{sec:partitions}

In this subsection, we fix conventions for integer partitions and recall several basic properties of partitions. 
For more details, see \cite[\S~I.1]{Mac98} and \cite[\S~5.D]{MOA11}.

Throughout this paper, a partition is a finite, possibly empty, ordered tuple of nonnegative integers.
Note that no monotonicity condition is imposed on its entries. Two partitions are equivalent if one is obtained from the other by permuting its entries, and sometimes the equality regarding partitions is understood in this sense.
For a partition \(\mathbf{d}=(d_{1},\dots,d_{l})\), we denote the weakly increasing and weakly decreasing rearrangements by
\begin{equation}
  \mathbf{d}^{\uparrow}=(d_{1}^{\uparrow},\cdots,d_{l}^{\uparrow}),\qquad\mathbf{d}^{\downarrow}=(d_{1}^{\downarrow},\cdots, d_{l}^{\downarrow}),
\end{equation}
respectively. The length and size of \(\mathbf{d}\) are defined by
\begin{equation}
  l(\mathbf{d}):=l,\qquad|\mathbf{d}|:=\sum_{i=1}^{l}d_{i},
\end{equation}
respectively. Zero entries are retained and counted in the length in this paper.
The empty partition has length and size zero. A partition is called positive if all its entries are positive. 
For \(s\in \mathbb{N}\), let \(\mathcal{P}_{s}\) denote the set of partitions of size \(s\), and let \(\mathcal{P}_{s}^{+}\) denote its subset of positive partitions. For \(l\in \mathbb{N}\), set
\begin{equation}
  \mathcal{P}_{s,l}:=\{ \mathbf{d}\in \mathcal{P}_{s}\mid l(\mathbf{d})=l \},\qquad \mathcal{P}_{s,l}^{+}=\mathcal{P}_{s}^{+}\cap \mathcal{P}_{s,l}.
\end{equation}
In tuple notation, \(a^{m}\) denotes \(m\) repetitions of the entry \(a\).
For partitions \(\mathbf{a}=(a_{1},\dots,a_{m})\) and \(\mathbf{b}=(b_{1},\dots,b_{n})\), we write 
\begin{equation}
  (\mathbf{a},\mathbf{b}):=(a_{1},\dots,a_{m},b_{1},\dots,b_{n})
\end{equation}
for their concatenation. We also denote this partition by \((a_{1},\dots,a_{m},\mathbf{b})\) or \((\mathbf{a},b_{1},\dots,b_{n})\). For \(n\in \mathbb{N}\), set \([n]:=\{ 1,\dots,n \}\). For \(\mathbf{d}=(d_{1},\dots,d_{l})\) and \(I\subseteq[l]\), let \(\mathbf{d}_{I}:=(d_{i})_{i\in I}\) denote the subpartition consisting of the entries indexed by \(I\).

Some common orders defined on \(\mathcal{P}_{s}^{+}\) are as follows. For details, see \cite{Mac98}.
\begin{definition}\label{def:order-Macdonald}
  Let \(\mathbf{a},\mathbf{b} \in \mathcal{P}_{s}^{+}\), and write their weakly decreasing rearrangements as
  \[
    \mathbf{a}^{\downarrow}=(a_{1}^{\downarrow},\dots,a_{m}^{\downarrow}),\qquad \mathbf{b}^{\downarrow}=(b_{1}^{\downarrow},\dots,b_{n}^{\downarrow}).
  \]
  For the comparisons below, we set \(a_{k}^{\downarrow}=0\) for \(k>m\) and \(b_{k}^{\downarrow}=0\) for \(k>n\).
  \begin{enumerate}[(1)]
    \item\textbf{Dominance order.} We write \(\mathbf{a}\trianglerighteq \mathbf{b}\) if 
    \begin{equation}\label{eq:major-ineq}
      \sum_{i=1}^{k}a_{i}^{\downarrow}\geq \sum_{i=1}^{k}b_{i}^{\downarrow},\qquad \forall k\geq1.
    \end{equation}
    \item \textbf{Reverse lexicographic order.} We write \(\mathbf{a}\succeq_{1}\mathbf{b}\) if 
    either \(\mathbf{a}=\mathbf{b}\) or \(a_{j}^{\downarrow}>b_{j}^{\downarrow}\) for the first \(j\) such that \(a_{j}^{\downarrow}\neq b_{j}^{\downarrow}\).
    \item \textbf{Reverse colexicographic order.} We write \(\mathbf{a}\succeq_{2}\mathbf{b}\) if either \(\mathbf{a}=\mathbf{b}\) or \(a_{s+1-j}^{\downarrow}<b_{s+1-j}^{\downarrow}\) for the first \(j\) such that \(a_{s+1-j}^{\downarrow}\neq b_{s+1-j}^{\downarrow}\).
  \end{enumerate}
  We write \(\mathbf{a}\triangleright \mathbf{b}\) if \(\mathbf{a}\trianglerighteq \mathbf{b}\) and \(\mathbf{a}\neq \mathbf{b}\). Similarly, for \(j=1,2\), we write \(\mathbf{a}\succ_{j}\mathbf{b}\) if \(\mathbf{a}\succeq_{j}\mathbf{b}\) and \(\mathbf{a}\neq \mathbf{b}\).
\end{definition}
\begin{remark}\label{rem:order-extension}
  The dominance order is only a partial order. The two reverse orders defined above are both total orders on \(\mathcal{P}_{s}^{+}\) extending the dominance order (see \cite[(1.10)]{Mac98}),
  i.e.,
  \[
    \mathbf{a}\trianglerighteq \mathbf{b}\implies \mathbf{a}\succeq_{j}\mathbf{b},\qquad j=1,2.
  \]
\end{remark}

\begin{remark}\label{rem:order-iso}
  Deleting all zero parts gives a bijection \(\iota\) from \(\mathcal{P}_{s,l}\) to \(\bigcup_{n\leq l}\mathcal{P}^{+}_{s,n}\subseteq \mathcal{P}_{s}^{+}\).
  Via this bijection, each order in Definition~\ref{def:order-Macdonald} induces an order on \(\mathcal{P}_{s,l}\), denoted by the same symbol; for example, \(\mathbf{a}\trianglerighteq\mathbf{b}\) means \(\iota(\mathbf{a})\trianglerighteq\iota(\mathbf{b})\).
  In particular, for \(\mathbf{a},\mathbf{b}\in\mathcal{P}_{s,l}\), we have \(\mathbf{a}\trianglerighteq\mathbf{b}\) if and only if \(\sum_{i=1}^{k}a_{i}^{\downarrow}\geq\sum_{i=1}^{k}b_{i}^{\downarrow}\) for all \(k\in[l]\).
\end{remark}

\subsection{Generalized Br\'ezin--Gross--Witten model and BKP hierarchy}
The generalized BGW model is a one-parameter deformation of the BGW model, with deformation parameter \(N \in \mathbb{C}\) \cite{MMS96}. Its partition function is defined by the normalized matrix integral
\begin{equation}
  \tau_{BGW}^{(N)}(\mathbf{t})=\frac{1}{\mathcal{Z}_{N}(\Lambda,\hbar)}\int[\dd \Phi]\exp\left(\mathrm{Tr}\left(\frac{\Lambda ^{2}\Phi+\Phi ^{-1}}{2\hbar}+(N-M)\log \Phi\right)\right),
  \ \ M\rightarrow\infty,
\end{equation}
where \(\Phi\) is an \(M\times M\) normal matrix, \(\Lambda\) is the external field, and the odd time variables \(\mathbf{t}=(t_{1},t_{3},t_{5},\dots)\) are given by the Miwa parametrization
\[
  t_{2k+1}= \frac{1}{2k+1}\mathrm{Tr}\Lambda ^{-2k-1}.
\]
The normalization factor \(\mathcal{Z}_{N}(\Lambda,\hbar)\) is chosen such that \(\tau_{BGW}^{(N)}(\mathbf{0})=1\). At \(N=0\), we recover the ordinary BGW tau-function.
Alexandrov \cite{Alex18} proved that the partition function of the generalized BGW model \(\tau_{BGW}^{(N)}(\mathbf{t})\) is determined by the normalization $\tau ^{(N)}_{BGW}(0)=1$
and the following Virasoro constraints
\begin{equation}\label{eq:Virasoro-gBGW}
  L_{m}^{(N)}\tau_{BGW}^{(N)}(\mathbf{t})=0,\quad \forall m\geq0,
\end{equation}
where the Virasoro operators \(\{ L_{m}^{(N)} \}_{m \in \mathbb{N}}\) are
\begin{equation}
  L_{m}^{(N)}:= \frac{1}{4}\sum_{\substack{a,b\geq0\\a+b=m-1}}\frac{ \partial ^{2} }{ \partial t_{2a+1}\partial t_{2b+1} } +\frac{1}{2}\sum_{c\geq0}(2c+1)t_{2c+1}\frac{ \partial  }{ \partial t_{2(c+m)+1} } -\frac{1}{2\hbar}\frac{ \partial  }{ \partial t_{2m+1} } + \frac{1-4N^{2}}{16}\delta_{m,0}.
\end{equation}
These operators satisfy the Virasoro commutation relation
\[
  \left[ L_{m}^{(N)},L_{n}^{(N)} \right] = (m-n)L_{m+n}^{(N)},\qquad m,n \geq0.
\]

Let \(\overline{\mathcal{M}}_{g,n}\) denote the moduli space of stable curves of genus \(g \) with \(n\) marked points, and let \(\psi_{i}\) denote the first Chern class of the cotangent line bundle at the \(i\)-th marked point. In the context of the generalized BGW model, Alexandrov--Norbury \cite{AN24} introduced the spin classes
\begin{equation}
  \Omega_{g,n+2h}^{(1^{n},0^{2h})}\in H^{4g-4+2n+2h}(\overline{\mathcal{M}}_{g,n+2h},\mathbb{Q}),
\end{equation}
and they proved that the generalized BGW correlators can be given by
\begin{equation}
  \langle \tau_{d_{1}}\cdots \tau_{d_{n}} \rangle_{g}^{\Omega,h}
  =\frac{1}{(2h)!}\int_{\overline{\mathcal{M}}_{g,n+2h}}\Omega_{g,n+2h}^{(1^{n},0^{2h})}\prod_{i=1}^n\psi_{i}^{d_i},\qquad h=|\mathbf{d}|-g+1.
\end{equation}
When \(h=0\), this construction specializes to the Theta class \(\Theta_{g,n}\) and the ordinary BGW correlators \cite{Nor23,CGG25}:
\begin{equation}
  \langle \tau_{d_{1}}\cdots\tau_{d_{n}} \rangle_{g}^{\Theta}= \int_{\overline{\mathcal{M}}_{g,n}}\Theta_{g,n}\prod_{i=1}^{n}\psi_{i}^{d_{i}}.
\end{equation}

In recent years, the partition function of the generalized BGW model \(\tau ^{(N)}_{BGW}(\mathbf{t})\) has been studied from the perspective of the BKP hierarchy.
It has been proved that \(\tau_{BGW}^{(N)}(\mathbf{t} /2)\) is a hypergeometric tau function of the BKP hierarchy and \(\tau_{BGW}^{(N)}(\mathbf{t})\) admits the following expansion in terms of Schur \(Q\)-functions \cite{Alex21b,Alex21a,LY22,Alex23}:
\begin{equation}\label{eq:gBGW-SchurQ}
  \tau_{BGW}^{(N)}(\mathbf{t})
  =\sum_{\lambda \in DP}\left(\frac{\hbar}{16}\right)^{|\lambda|}2^{-l(\lambda)}\left(\prod_{i=1}^{l(\lambda)}\prod_{j=1}^{\lambda_{i}}((2j-1)^{2}-4N^{2})\right)Q_{\lambda}(\delta_{k,1})Q_{\lambda}\left(\mathbf{t}\right),
\end{equation}
where \(DP\) denotes the set of strict partitions \(\lambda=(\lambda_{1}>\lambda_{2}>\cdots>\lambda_{l})\) with positive parts, and \(Q_{\lambda}\left(\mathbf{t}\right)\) denotes the Schur \(Q\)-function associated with \(\lambda\). In particular,
\[
  Q_{\lambda}(\delta_{k,1})= \frac{2^{|\lambda|}}{\prod_{i=1}^{l(\lambda)}\lambda_{i}!}
  \cdot \prod_{i<j} \frac{\lambda_{i}-\lambda_{j}}{\lambda_{i}+\lambda_{j}}.
\]

The BKP hierarchy is the type-B analogue of the Kadomtsev--Petviashvili (KP) hierarchy introduced by the Kyoto School \cite{DJKM82,JM83}. A BKP tau function \(\tau\) normalized by \(\tau(\mathbf{0})=1\) is uniquely determined by its BKP affine coordinates on the big cell of the isotropic Sato Grassmannian \cite{HB21,WY22}.
In particular,
the connected bosonic \(n\)-point functions provide formulas for recovering correlators from a tau function of the BKP hierarchy using the affine coordinates; see \cite{WY22,ZZ26}.
For the generalized BGW model,
the corresponding BKP affine coordinates can be read from the Schur \(Q\)-function expansion \eqref{eq:gBGW-SchurQ} of \(\tau ^{(N)}_{BGW}(\mathbf{t})\).

We first recall several results on the type B affine coordinates of the generalized BGW model \cite{WY22,WYZ25}.
The BKP affine coordinates of \(\tau_{BGW}^{(N)}(\mathbf{t} /2)\) are given by (see Proposition 3.1 in \cite{WYZ25})
\begin{align}
  a_{0,n}^{(N)}&=-a_{n,0}^{(N)}= \frac{\hbar^{n}}{2^{3n+1}n!}\prod_{k=1}^{n}\left( (2k-1)^{2}-4N^{2} \right) , &n>0,\\ 
  a_{m,n}^{(N)}&= \frac{\hbar^{m+n}}{2^{3m+3n+2}m!n!}\frac{n-m}{m+n}\prod_{j=1}^{m}\left( (2j-1)^{2}-4N^{2} \right)\prod_{k=1}^{n}\left( (2k-1)^{2}-4N^{2} \right)  ,&m,n>0.
\end{align}
Following \cite{WY22,GJY25},
denote the following two functions
\begin{align}\label{eqn:phi1phi2}
	\begin{split}
  \Phi_{1}^{(N)}(z)&=1+\sum_{k=1}^{\infty} \frac{(-\hbar)^{k}}{8^{k}k!}\prod_{i=1}^{k}\left( 4N^{2}-(2i-1)^{2} \right) z^{-k},\\ 
  \Phi_{2}^{(N)}(z)&=z+\sum_{k=1}^{\infty} \frac{(-\hbar)^{k}}{8^{k}k!}\prod_{i=1}^{k}(4(1-N)^{2}-(2i-1)^{2})z^{1-k}.
  \end{split}
\end{align}
 Then the generating series of the BKP affine coordinates can be given by
\begin{align}
  A^{(N)}(w,z)&=\sum_{n,m>0}(-1)^{m+n+1}a_{n,m}^{(N)}w^{-n}z^{-m}-\sum_{n>0} \frac{(-1)^{n}}{2}a_{n,0}^{(N)}(w^{-n}-z^{-n}) ,\\
  \hat{A}^{(N)}(w,z)&=A^{(N)}(w,z)-\frac{1}{4}-\frac{1}{2}\sum_{i=1}^{\infty}(-1)^{i}w^{-i}z^{i},\\ 
  &= \iota_{w,z}\frac{\Phi_{1}^{(N)}(-z)\Phi_{2}^{(N)}(-w)-\Phi_{1}^{(N)}(-w)\Phi_{2}^{(N)}(-z)}{4(w+z)},
\end{align}
where the notation $\iota_{w,z}$ means we expand the function in the region $|w|>|z|$.
Define \(\xi(z_{i},z_{j})\) by 
\[
  \xi(z_{i},z_{j})=\begin{cases}
    \hat{A}^{(N)}(z_{i},z_{j}),&i<j,\\ 
    A^{(N)}(z_{i},z_{j}), &i=j, \\
    -\hat{A}^{(N )}(z_{j},z_{i}),&i>j.
  \end{cases}
\]
Then the connected \(n\)-point function of the generalized BGW model is given by (see \cite{WY22,ZZ26})
\begin{align}
	\begin{split}
  &\sum_{i_{1},\dots,i_{n}>0\text{ odd}}\frac{ \partial ^{n}\log\tau_{BGW}^{(N)}(\mathbf{t}/2) }{ \partial t_{i_{1}}\cdots \partial t_{i_{n}} }\Bigg|_{\mathbf{t}=0} z_{1}^{-i_{1}}\cdots z_{n}^{-i_{n}}\\
  =&-2^{n-1}\bigg( \sum_{\sigma:n-\text{cycles}}\prod_{i=1}^{n}\xi(z_{i},-z_{\sigma(i)}) \bigg)_{\text{odd}} 
  -\delta_{n,2}\sum_{k=0}^{\infty} \frac{2k+1}{2}z_{1}^{-2k-1}z_{2}^{2k+1},
  \end{split} \label{eq:correction-term}
\end{align}
where \((-)_{\text{odd}}\) retains only those Laurent monomials in which each variable \(z_{i}\) appears with an odd exponent. Consequently, 
\begin{equation}\label{eq:gBGW-n-point}
  \frac{ \partial ^{n}\log \tau_{BGW}^{(N)}(\mathbf{t}) }{ \partial t_{2d_{1}+1}\cdots \partial t_{2d_{n}+1} }\Bigg|_{\mathbf{t}=0}
  =-2^{2n-1}\left[ z_{1}^{-2d_{1}-1}\cdots z_{n}^{-2d_{n}-1} \right] \bigg( \sum_{\sigma:n-\text{cycles}}\prod_{i=1}^{n}\xi(z_{i},-z_{\sigma(i)}) \bigg),
\end{equation}
since the correction term $\delta_{n,2}\sum_{k=0}^{\infty} \frac{2k+1}{2}z_{1}^{-2k-1}z_{2}^{2k+1}$ makes no contribution to the indicated coefficient. Here \([z_{1}^{k_{1}}\cdots z_{n}^{k_{n}}]F\) denotes the coefficient of \(z_{1}^{k_{1}}\cdots z_{n}^{k_{n}}\) in the formal Laurent expansion of \(F\). 

\section{Concavity of the generalized BGW correlators}\label{sec:concavity}
In this section, we prove the first of our main results, Theorem~\ref{thm:intro-concavity}, which establishes the strict concavity of the normalized generalized BGW correlators.
For convenience,
in the following of this paper,
we will set \(\hbar=1\). 

\subsection{Generalized BGW coefficients and their second differences}\label{sec:generalized-bgw-coefficients}
In this subsection, we introduce several basic concepts used throughout this paper, including the generalized BGW coefficients and their normalization, as well as the concavity of a function defined on \(\mathcal{P}_{s}\) of a given size \(s\). We also prove that these coefficients are positive.

For a nonempty partition \(\mathbf{d}=(d_{1},\dots,d_{l})\), define
\begin{align}
  B^{(N)}(\mathbf{d}) & =\frac{ \partial ^{l}\log \tau_{BGW}^{(N)}(\mathbf{t}) }{ \partial t_{2d_{1}+1}\cdots \partial t_{2d_{l}+1} }\Bigg|_{\mathbf{t}=0}, \\
  \chi(\mathbf{d}) & =2|\mathbf{d}|+l(\mathbf{d}).
\end{align}
In particular,
we denote \(\chi(\emptyset)=0\).
\begin{proposition}\label{prp:Bhd-positive}
  For every nonempty partition \(\mathbf{d}\), \(B^{(N)}(\mathbf{d})\) is a polynomial in \(-N^{2}\) of degree \(|\mathbf{d}|+1\), all of whose coefficients are positive.
\end{proposition}

\begin{proof}
  We prove this proposition by induction on \(\chi(\mathbf{d})=2|\mathbf{d}|+l(\mathbf{d})\). For \(m\geq0\) and \(\mathbf{d}=(d_{1},\dots,d_{l})\), the Virasoro constraints \eqref{eq:Virasoro-gBGW} for \(\tau_{BGW}^{(N)}(\mathbf{t})\) are equivalent to
  \begin{equation}\label{eq:Virasoro-B}
    \begin{aligned}
      B^{(N)}(m,\mathbf{d}) & =\frac{1}{2}\sum_{\substack{a,b\geq0\\a+b=m-1}}\bigg(B^{(N)}(a,b,\mathbf{d})+\sum_{I\sqcup J=[l(\mathbf{d})]}B^{(N)}(a,\mathbf{d}_{I})B^{(N)}(b,\mathbf{d}_{J})\bigg) \\
      & +\sum_{k=1}^{l}(2d_{k}+1)B^{(N)}(d_{1},\dots,d_{k-1},d_{k}+m,d_{k+1},\dots,d_{l})+ \frac{1-4N^{2}}{8}\delta_{m,0}\delta_{l,0}.
    \end{aligned}
  \end{equation}
  Sums over \(I \sqcup J=[l(\mathbf{d})]\) run over ordered pairs of complementary subsets, including empty subsets. Taking \(m=0\) and \(\mathbf{d}=\emptyset\), one has
  \begin{equation}
    B^{(N)}(0)= \frac{1-4N^{2}}{8}.
  \end{equation}
  This proves the base case \(\mathbf{d}=(0)\), for which \(\chi(\mathbf{d})=1\). 
  
  Assume that the proposition holds for every nonempty partition \(\mathbf{d}\) with \(\chi(\mathbf{d})\leq M\). Suppose that \(\chi(m,\mathbf{d})=M+1\). 
  Then the values of \(\chi\) of all partitions appeared on the right-hand side of \eqref{eq:Virasoro-B} are at most \(M\). By the induction hypothesis, every term on the right-hand side is a polynomial in $-N^2$,
  has nonnegative coefficients and is of degree at most \(m+|\mathbf{d}|+1\).
  Moreover,
  if \(m\geq1\), each product term in the first row of \eqref{eq:Virasoro-B} has degree exactly \(m+|\mathbf{d}|+1\) with all coefficients positive,
  and if \(m=0\), then \(l\geq1\), the second row of \eqref{eq:Virasoro-B} has the same property.
  Thus $B^{(N)}(m,\mathbf{d})$ must be a polynomial in $-N^2$ with positive coefficients and degree $m+|\mathbf{d}|+1$.
  This completes the proof.
\end{proof}

\begin{definition}\label{def:gBGW-coefficient}
  For a nonempty partition \(\mathbf{d}\), define the generalized BGW coefficients \(B_{h}(\mathbf{d})\) by
  \begin{equation}
    B^{(N)}(\mathbf{d})=\sum_{h=0}^{|\mathbf{d}|+1}B_{h}(\mathbf{d})(-N^{2})^{h}.
  \end{equation}
\end{definition}
For \(\mathbf{d}=(d_{1},\dots,d_{l})\), the generalized BGW coefficient \(B_{h}(\mathbf{d})\) is related to the generalized BGW correlator \(\langle \tau_{d_{1}}\cdots\tau_{d_{l}} \rangle_{|\mathbf{d}|+1-h}^{\Omega}\) by
\begin{equation*}
  B_{h}(\mathbf{d}) =\prod_{i=1}^{l}(2d_{i}+1)!!\langle \tau_{d_{1}}\cdots\tau_{d_{l}} \rangle_{|\mathbf{d}|+1-h}^{\Omega,h}.
\end{equation*}
In particular,
it is related to the normalized generalized BGW correlator \(C_{h}(\mathbf{d})\) by 
  \begin{equation}\label{eq:def-C}
    C_{h}(\mathbf{d})=\kappa_{\mathbf{d}}^{-1}B_{h}(\mathbf{d}),
  \end{equation}
  where the normalization factor is given by $\kappa_{\mathbf{d}}= \frac{(2|\mathbf{d}|+l(\mathbf{d})-1)!}{2^{2|\mathbf{d}|+1}}.$

\begin{definition}
  For a given size \(s\), a function \(f:\mathcal{P}_{s}\to \mathbb{R}\) is said to be strictly concave if, for every partition \(\mathbf{d}=(d_{1},\dots,d_{l})\in \mathcal{P}_{s}\) and all distinct \(i,j \in[l]\) satisfying \(d_{i},d_{j}>0\), the following inequality holds:
  \begin{equation}
    f(\mathbf{d}+e_{i}-e_{j})+f(\mathbf{d}+e_{j}-e_{i})<2f(\mathbf{d}).
  \end{equation}
\end{definition}
For a partition \(\mathbf{d}=(d_{1},\dots,d_{l})\) and distinct indices \(i,j \in [l]\) with \(d_{i},d_{j}>0\), define the second difference \(K_{i,j}^{(N)}(\mathbf{d})\) of the function \(B^{(N)}\) by
\begin{equation}\label{eqn:def KNij}
  K_{i,j}^{(N)}(\mathbf{d})=B^{(N)}(\mathbf{d}+e_{i}-e_{j})+B^{(N)}(\mathbf{d}+e_{j}-e_{i})-2B^{(N)}(\mathbf{d}).
\end{equation}
Denote the expansion of \(K_{i,j}^{(N)}(\mathbf{d})\) in \(-N^{2}\) by
\begin{equation}\label{def:KijN as Kh}
  K_{i,j}^{(N)}(\mathbf{d})=\sum_{h=0}^{|\mathbf{d}|+1}K_{i,j;h}(\mathbf{d})(-N^{2})^{h}.
\end{equation}
Throughout this paper, the notation \(K_{i,j}^{(N)}(\mathbf{d})\) and \(K_{i,j;h}(\mathbf{d})\) is used only for distinct indices \(i,j\in[l(\mathbf{d})]\) satisfying \(d_{i},d_{j}>0\).
For a general partition \(\mathbf{d}\), following Definition \ref{def:gBGW-coefficient}, equations \eqref{eqn:def KNij}, \eqref{def:KijN as Kh},
and comparing coefficients of \((-N^{2})^{h}\) in the definition of \(K_{i,j}^{(N)}\) give
\begin{equation}\label{eq:def-K-h}
  K_{i,j;h}(\mathbf{d})=B_{h}(\mathbf{d}+e_{i}-e_{j})+B_{h}(\mathbf{d}+e_{j}-e_{i})-2B_{h}(\mathbf{d}).
\end{equation}
By equation \eqref{eq:def-C},
the generalized BGW correlators \(C_{h}(\mathbf{d})\) differ from the generalized BGW coefficients \(B_{h}(\mathbf{d})\) by a global factor $\kappa_{\mathbf{d}}= \frac{(2|\mathbf{d}|+l(\mathbf{d})-1)!}{2^{2|\mathbf{d}|+1}}$,
which depends only on the size and the length of the partition $\mathbf{d}$.
Thus, the strict concavity of the generalized BGW correlators \(C_{h}\) is equivalent to the strict concavity of \(B_{h}\), which is in turn equivalent to \(K_{i,j;h}(\mathbf{d})<0\).

\subsection{Computations of shifted second differences}
\label{sec:shifted}
Motivated by the fact that the time translation preserves the BKP integrability,
we define the normalized shifted tau function by
\begin{equation}
  \tau_{BGW}^{(N)}(\mathbf{t}; \mathbf{s}):= \frac{\tau_{BGW}^{(N)}(\mathbf{s}+\mathbf{t})}{\tau_{BGW}^{(N)}(\mathbf{s})},
\end{equation}
where $\mathbf{s}=(s_1,s_3,\dots)$.
For \(\mathbf{d}=(d_{1},\dots,d_{l})\) with \(l\geq1\), define the shifted generalized BGW coefficients by
\begin{equation}\label{eqn:def shif BGW}
  B^{(N)}(\mathbf{d}; \mathbf{s}):= \frac{\partial ^{l}\log \tau_{BGW}^{(N)}(\mathbf{t}; \mathbf{s})}{\partial  t_{2d_{1}+1}\cdots \partial t_{2d_{l}+1}}\Bigg|_{\mathbf{t}=0}.
\end{equation}
and for distinct indices \(i,j\in[l]\) with \(d_{i},d_{j}\in \mathbb{N}_{+}\), define the corresponding shifted second difference \(K^{(N)}_{i,j}(\mathbf{d}; \mathbf{s})\) by 
\[
  K^{(N)}_{i,j}(\mathbf{d};\mathbf{s}):=B^{(N)}(\mathbf{d}+e_{i}-e_{j}; \mathbf{s})+B^{(N)}(\mathbf{d}+e_{j}-e_{i}; \mathbf{s})-2B^{(N)}(\mathbf{d}; \mathbf{s}).
\]

For any partition \(\mathbf{a}\), write
\begin{equation}\label{eq:differential-partition}
  \partial_{\mathbf{a}}=\prod_{i=1}^{l(\mathbf{a})}\frac{ \partial  }{ \partial s_{2a_{i}+1} }.
\end{equation}
Then the definition \eqref{eqn:def shif BGW} of the shifted generalized BGW coefficients gives
\begin{equation}\label{eq:def-Bs}
  B^{(N)}(\mathbf{d}; \mathbf{s})=\partial_{\mathbf{d}}\log \tau_{BGW}^{(N)}(\mathbf{s}).
\end{equation}
For the case of $l(\mathbf{d})=2$ with $\mathbf{d}=(d_1,d_2)$, we use the following shorthand notation
\begin{equation}
  K^{(N)}(d_{1},d_{2})=K_{1,2}^{(N)}(d_{1},d_{2}),\qquad K_{h}(d_{1},d_{2})=K_{1,2;h}(d_{1},d_{2}),
\end{equation}
and the shifted version $K^{(N)}(d_{1},d_{2};\mathbf{s})=K_{1,2}^{(N)}(d_{1},d_{2}; \mathbf{s})$.
By \eqref{eq:def-Bs}, for distinct indices \(i,j\in[l]\) with \(d_{i},d_{j}\geq1\), we obtain
\begin{equation}\label{eq:K-ShiftedK}
  K_{i,j}^{(N)}(\mathbf{d})= \partial_{\mathbf{d}_{[l]\setminus \{ i,j \}}}K^{(N)}(d_{i},d_{j};\mathbf{s})\Big|_{\mathbf{s}=0}.
\end{equation}
Thus, to compute the second differences \(K_{i,j}^{(N)}(\mathbf{d}), K_{i,j;h}(\mathbf{d})\) of the generalized BGW coefficients, it suffices to compute the shifted two-point second difference \(K^{(N)}(x,y; \mathbf{s})\). 

Motivated by \eqref{eqn:phi1phi2},
define the following two shifted functions
\begin{align}
  \Phi_{1}^{(N)}(z; \mathbf{s})&:=\frac{\tau_{BGW}^{(N)}(\mathbf{s}+[z^{-1}]_{\text{odd}})}{\tau_{BGW}^{(N)}(\mathbf{s})}, \\
  \Phi_{2}^{(N)}(z; \mathbf{s})&:=\left( z-\frac{1}{2}-\frac{ \partial  }{ \partial s_{1} }  \right) \Phi_{1}^{(N)}(z; \mathbf{s}),
\end{align}
where we use the notation $[z]_{\text{odd}}=\big(z,\frac{z^{3}}{3},\frac{z^{5}}{5},\dots\big).$
Then the following explicit expression for the shifted two-point second difference \(K^{(N)}(x,y; \mathbf{s})\) holds.
\begin{lemma}\label{lem:KN as xi}
  For \(x,y\in \mathbb{N}_{+}\), the second difference \(K^{(N)}(x,y; \mathbf{s})\) is given by 
  \begin{equation}\label{eqn:KN as xi}
    K^{(N)}(x,y; \mathbf{s})=-8[z_{1}^{-2x-1}z_{2}^{-2y-1}]\left( \frac{z_{1}}{z_{2}}-\frac{z_{2}}{z_{1}} \right)^{2}\left( \xi(z_{1},-z_{2}; \mathbf{s})\xi(z_{2},-z_{1}; \mathbf{s}) \right),
  \end{equation}
  where $\xi(z_{1},-z_{2}; \mathbf{s})
  =\hat{A}^{(N)}(z_{1},-z_{2}; \mathbf{s}),\ \ 
  \xi(z_{2},-z_{1}; \mathbf{s})
  =-\hat{A}^{(N)}(-z_{1},z_{2}; \mathbf{s})$,
  and
  \begin{equation}\label{eq:Ahats}
    \hat{A}^{(N)}(w,z; \mathbf{s})= \frac{\Phi_{1}^{(N)}(-z; \mathbf{s})\Phi_{2}^{(N)}(-w; \mathbf{s})-\Phi_{1}^{(N)}(-w; \mathbf{s})\Phi_{2}^{(N)}(-z; \mathbf{s})}{4(w+z)}.
  \end{equation}
\end{lemma}
\begin{proof}
  Applying \cite[Proposition~3.5]{WY22} to the BKP tau-function \(\tau_{BGW}^{(N)}(\mathbf{t}/ 2; \mathbf{s})\), we obtain
  \begin{equation}\label{eq:Ahat-tau}
    \hat{A}^{(N)}(w,z;\mathbf{s})  = -\frac{w-z}{4(w+z)}\tau_{BGW}^{(N)}(-[w^{-1}]_{\text{odd}}-[z^{-1}]_{\text{odd}}; \mathbf{s}).
  \end{equation}
  Note that the partition function \(\tau_{BGW}^{(N)}(\mathbf{t}; \mathbf{s})\) is also a tau-function of the KP hierarchy with respect to the time variables $(t_1,0,t_3,0,t_5,\dots)$.
  The differential Fay identity for the KP hierarchy therefore gives (see \cite{AvM92})
  \begin{align*}
    & \tau_{BGW}^{(N)}(\mathbf{s}-[z^{-1}]_{\text{odd}})\frac{ \partial  }{ \partial s_{1} } \tau_{BGW}^{(N)}(\mathbf{s}-[w^{-1}]_{\text{odd}})-\tau_{BGW}^{(N)}(\mathbf{s}-[w^{-1}]_{\text{odd}})\frac{ \partial  }{ \partial s_{1} } \tau_{BGW}^{(N)}(\mathbf{s}-[z^{-1}]_{\text{odd}}) \\
    = & (w-z)(\tau_{BGW}^{(N)}(\mathbf{s}-[z^{-1}]_{\text{odd}}-[w^{-1}]_{\text{odd}})\tau_{BGW}^{(N)}(\mathbf{s})-\tau_{BGW}^{(N)}(\mathbf{s}-[z^{-1}]_{\text{odd}})\tau_{BGW}^{(N)}(\mathbf{s}-[w^{-1}]_{\text{odd}})).
  \end{align*}
By the definitions of \(\Phi_{1}^{(N)}(z; \mathbf{s})\) and \(\Phi_{2}^{(N)}(z; \mathbf{s})\), we have
  \begin{align*}
    \frac{\partial_{(0)} \tau_{BGW}^{(N)}(\mathbf{s}+[z^{-1}]_{\text{odd}}) }
    {\tau_{BGW}^{(N)}(\mathbf{s})}
    & = \bigg(z- \frac{1}{2}+\frac{ \partial \log \tau_{BGW}^{(N)}(\mathbf{s}) }{ \partial s_{1} } \bigg)\Phi_{1}^{(N)}(z; \mathbf{s})-\Phi_{2}^{(N)}(z; \mathbf{s}).
  \end{align*}
  Dividing the differential Fay identity by \(\tau_{BGW}^{(N)}(\mathbf{s})^{2}\) and using the above relation with \(z\) replaced by \(-z\) and \(-w\), we obtain 
  \begin{equation}\label{eq:tau-Phi}
    (z-w)\tau_{BGW}^{(N)}(-[z^{-1}]_{\text{odd}}-[w^{-1}]_{\text{odd}};\mathbf{s})=\Phi_{1}^{(N)}(-z; \mathbf{s})\Phi_{2}^{(N)}(-w; \mathbf{s})-\Phi_{1}^{(N)}(-w; \mathbf{s})\Phi_{2}^{(N)}(-z; \mathbf{s}).
  \end{equation}
  Substituting \eqref{eq:tau-Phi} into \eqref{eq:Ahat-tau} yields
  \begin{equation}
    \hat{A}^{(N)}(w,z;\mathbf{s}) = \frac{\Phi_{1}^{(N)}(-z; \mathbf{s})\Phi_{2}^{(N)}(-w; \mathbf{s})-\Phi_{1}^{(N)}(-w; \mathbf{s})\Phi_{2}^{(N)}(-z; \mathbf{s})}{4(w+z)}.
  \end{equation}
  Applying the formula \eqref{eq:gBGW-n-point} to \(\tau_{BGW}^{(N)}(\mathbf{t}/2; \mathbf{s})\) with \(n=2\), we obtain a formula for the two-point generalized BGW coefficients \(B^{(N)}(x,y; \mathbf{s})\). Substituting the above expression for \(\hat{A}^{(N)}(w,z;\mathbf{s})\) and taking the second difference in \(x\) and \(y\) gives the desired formula \eqref{eqn:KN as xi}.
\end{proof}

Define the shifted auxiliary functions by
\begin{equation}
  \Phi_{p,q}^{(N)}(z; \mathbf{s}) := \frac{\Phi_{p}^{(N)}(z; \mathbf{s})\Phi_{q}^{(N)}(-z; \mathbf{s})}{z},\quad p,q\in\{ 1,2 \}.
\end{equation}
The following lemma expresses these shifted auxiliary functions in terms of the shifted generalized BGW coefficients \(B^{(N)}(\mathbf{d}; \mathbf{s})\). 
\begin{lemma}\label{lem:Phis-Bs}
  The shifted auxiliary functions \(\Phi_{p,q}^{(N)}(z; \mathbf{s})\) are given explicitly by the following expressions:
  \begin{align*}
    \Phi_{1,1}^{(N)}(z; \mathbf{s}) & = \frac{1}{z}+\sum_{k=1}^{\infty}B^{(N)}(0,k-1; \mathbf{s})z^{-2k-1}, \\
    \Phi_{1,2}^{(N)}(z; \mathbf{s}) & =  -1- \frac{1}{2z}-\frac{1}{2}\sum_{k=1}^{\infty}\left(B^{(N)}(0,0,k-1; \mathbf{s})+B^{(N)}(0,k-1; \mathbf{s})\right)z^{-2k-1}, \\
    \Phi_{2,2}^{(N)}(z; \mathbf{s}) & =-z+\left( \frac{1}{4}+u-B^{(N)}(0^{2}; \mathbf{s}) \right) \frac{1}{z}+\sum_{k=1}^{\infty}\left(\frac{1}{2}B^{(N)}(0^{3},k-1;\mathbf{s})\right.\\
    &\quad\quad \left.+\frac{1}{2}B^{(N)}(0^{2},k-1; \mathbf{s})+\left(\frac{1}{4}+u\right)B^{(N)}(0,k-1; \mathbf{s})-B^{(N)}(0,k; \mathbf{s})\right)z^{-2k-1},
  \end{align*}
  where $u=2\cdot \partial_{(0,0)} \log \tau_{BGW}^{(N)}(\mathbf{s})$.
\end{lemma}
\begin{proof}
  \textbf{Computation of \(\Phi_{1,1}^{(N)}(z; \mathbf{s})\).} By definition, \(\Phi_{1,1}^{(N)}(z; \mathbf{s})\) is odd in \(z\). For \(k\in \mathbb{N}\), the coefficient of \(z^{-2k-1}\) in \(\Phi_{1,1}^{(N)}(z; \mathbf{s})\) is given by
\begin{equation}\label{eq:coeff-Phi11}
[z^{-2k-1}]\Phi_{1,1}^{(N)}(z; \mathbf{s}) =\frac{1}{\tau_{BGW}^{(N)}(\mathbf{s})^{2}}\operatorname{Res}_{z=\infty}z^{2k-1}\tau_{BGW}^{(N)}(\mathbf{s}-[z^{-1}]_{\text{odd}})\tau_{BGW}^{(N)}(\mathbf{s}+[z^{-1}]_{\text{odd}})\dd z,
\end{equation}
    where \(\operatorname{Res}_{z=\infty}f(z)\,\dd z\) denotes the coefficient of \(z^{-1}\) in \(f(z)\).

    For \(k=0\), we have
\begin{equation}\label{eq:coeff-Phi11-0}
  [z^{-1}]\Phi_{1,1}^{(N)}(z; \mathbf{s})=1.
\end{equation}
    For \(k\geq1\), recall that the partition function of the generalized BGW model \(\tau_{BGW}^{(N)}(\mathbf{s})\) satisfies the bilinear identity for the KdV hierarchy (see \cite{MMS96,Alex18}):
  \begin{equation}
    \operatorname{Res}_{z=\infty}\exp\left(\sum_{j=0}^{\infty}(s_{2j+1}-s_{2j+1}')z^{2j+1}\right)\tau_{BGW}^{(N)}(\mathbf{s}-[z^{-1}]_{\text{odd}})\tau_{BGW}^{(N)}(\mathbf{s}'+[z^{-1}]_{\text{odd}})\dd z=0.
  \end{equation}
Differentiating both sides of the preceding equation with respect to \(s_{2k-1}\) yields
  \begin{equation}\label{eq:KdV-dt2k-1}
    \begin{aligned}
      0  =\operatorname{Res}_{z=\infty}&\exp\bigg(\sum_{j=0}^{\infty}(s_{2j+1}-s_{2j+1}')z^{2j+1}\bigg)\\
      &\left(z^{2k-1}\tau_{BGW}^{(N)}(\mathbf{s}-[z^{-1}]_{\text{odd}})
      +\frac{ \partial \tau_{BGW}^{(N)}(\mathbf{s}-[z^{-1}]_{\text{odd}}) }{ \partial s_{2k-1} } \right)\tau_{BGW}^{(N)}(\mathbf{s}'+[z^{-1}]_{\text{odd}})\dd z.
    \end{aligned}
  \end{equation}
   By \eqref{eq:coeff-Phi11}, setting \(\mathbf{s}=\mathbf{s}'\) in \eqref{eq:KdV-dt2k-1} gives
\begin{align*}
[z^{-2k-1}]\Phi_{1,1}^{(N)}(z; \mathbf{s}) =-\frac{1}{\tau_{BGW}^{(N)}(\mathbf{s})^{2}}\operatorname{Res}_{z=\infty} \tau_{BGW}^{(N)}(\mathbf{s}+[z^{-1}]_{\text{odd}})\frac{ \partial \tau_{BGW}^{(N)}(\mathbf{s}-[z^{-1}]_{\text{odd}}) }{ \partial s_{2k-1} } \dd z .
\end{align*}
By direct computations,
the function appearing in the right hand side of above equation is equal to
\begin{align*}
  & \tau_{BGW}^{(N)}(\mathbf{s}+[z^{-1}]_{\text{odd}})\frac{ \partial \tau_{BGW}^{(N)}(\mathbf{s}-[z^{-1}]_{\text{odd}}) }{ \partial s_{2k-1} } \\
  = & \left( \tau_{BGW}^{(N)}(\mathbf{s})+\frac{ \partial \tau_{BGW}^{(N)}(\mathbf{s}) }{ \partial s_{1} } \frac{1}{z}+O\left(\frac{1}{z^{2}}\right) \right) \left( \frac{ \partial \tau_{BGW}^{(N)}(\mathbf{s}) }{ \partial s_{2k-1} } -\frac{ \partial ^{2}\tau_{BGW}^{(N)}(\mathbf{s}) }{ \partial s_{1}\partial s_{2k-1} } \frac{1}{z}+O\left(\frac{1}{z^{2}}\right) \right) \\
  = & \tau_{BGW}^{(N)}(\mathbf{s})\frac{ \partial \tau_{BGW}^{(N)}(\mathbf{s}) }{ \partial s_{2k-1} }-\tau_{BGW}^{(N)}(\mathbf{s})^{2}\frac{ \partial ^{2}\log \tau_{BGW}^{(N)}(\mathbf{s}) }{ \partial s_{1}\partial s_{2k-1} } \frac{1}{z}+O\left(\frac{1}{z^{2}}\right).
\end{align*}
Thus, we obtain
\begin{equation}\label{eq:coeff-Phi11-1}
[z^{-2k-1}]\Phi_{1,1}^{(N)}(z; \mathbf{s})=\frac{ \partial ^{2}\log \tau_{BGW}^{(N)}(\mathbf{s}) }{ \partial s_{1}\partial s_{2k-1} }=B^{(N)}(0,k-1; \mathbf{s}).
\end{equation}
The computation of \(\Phi_{1,1}^{(N)}(z;\mathbf{s})\) is completed by combining \eqref{eq:coeff-Phi11-0} and \eqref{eq:coeff-Phi11-1}.

  \textbf{Computation of \(\Phi_{1,2}^{(N)}(z; \mathbf{s})\).} By the definition of \(\Phi_{2}^{(N)}(z; \mathbf{s})\), we have
\begin{align*}
\Phi_{1,2}^{(N)}(z; \mathbf{s}) & =\frac{1}{z}\left( \Phi_{1}^{(N)}(z; \mathbf{s})\left(-z-\frac{1}{2}- \frac{ \partial  }{ \partial s_{1} }  \right)\Phi_{1}^{(N)}(-z; \mathbf{s}) \right) \\
 & =-\left(z+\frac{1}{2}\right)\Phi_{1,1}^{(N)}(z; \mathbf{s})-\frac{1}{z}\Phi_{1}^{(N)}(z; \mathbf{s})\frac{ \partial  }{ \partial s_{1} } \Phi_{1}^{(N)}(-z; \mathbf{s}).
\end{align*}
Since \(\Phi_{1,1}^{(N)}(z; \mathbf{s})\) is odd, replacing \(z\) by \(-z\) gives
\[
\Phi_{1,2}^{(N)}(-z; \mathbf{s})=\left(\frac{1}{2}-z\right)\Phi_{1,1}^{(N)}(z; \mathbf{s})+\frac{1}{z}\Phi_{1}^{(N)}(-z; \mathbf{s})\frac{ \partial  }{ \partial s_{1} } \Phi_{1}^{(N)}(z; \mathbf{s}).
\]
  Hence, we have
\begin{equation}\label{eq:Phi12-1}
  \Phi_{1,2}^{(N)}(-z; \mathbf{s})-\Phi_{1,2}^{(N)}(z; \mathbf{s})=\Phi_{1,1}^{(N)}(z; \mathbf{s})+\frac{ \partial  }{ \partial s_{1} } \Phi_{1,1}^{(N)}(z; \mathbf{s}).
\end{equation}
  Setting \(w=-z\) in \eqref{eq:tau-Phi} gives
    \begin{equation}\label{eq:Phi12-2}
  	\Phi_{1,2}^{(N)}(-z; \mathbf{s})+\Phi_{1,2}^{(N)}(z; \mathbf{s})=-2.
  \end{equation}
  Then we have
  \begin{align*}
    \Phi_{1,2}^{(N)}(z; \mathbf{s})& =-1-\frac{1}{2}\left( 1+\frac{ \partial  }{ \partial s_{1} } \right)  \Phi_{1,1}^{(N)}(z;\mathbf{s}) \\
    & = -1- \frac{1}{2z}-\frac{1}{2}\sum_{k=1}^{\infty}\left(B^{(N)}(0,k-1; \mathbf{s})+B^{(N)}(0^{2},k-1; \mathbf{s})\right)z^{-2k-1}.
  \end{align*}
  
  \textbf{Computation of \(\Phi_{2,2}^{(N)}(z; \mathbf{s})\).} By the definition of \(\Phi_{2}^{(N)}(z; \mathbf{s})\), we have
  \begin{equation}\label{eq:Phi22-1}
    \begin{aligned}
      \Phi_{2,2}^{(N)}(z; \mathbf{s})
      =& \frac{1}{z}\left(z-\frac{1}{2}-\frac{ \partial  }{ \partial s_{1} } \right)\Phi_{1}^{(N)}(z; \mathbf{s})\cdot\left(-z-\frac{1}{2}-\frac{ \partial  }{ \partial s_{1} } \right)\Phi_{1}^{(N)}(-z; \mathbf{s}) \\
      =&\left(\frac{1}{4}-z^{2}\right)\Phi_{1,1}^{(N)}(z; \mathbf{s})+\frac{1}{2}\frac{ \partial  }{ \partial s_{1} } \Phi_{1,1}^{(N)}(z; \mathbf{s})+\Phi_{1}^{(N)}(-z; \mathbf{s})\frac{ \partial  }{ \partial s_{1} } \Phi_{1}^{(N)}(z; \mathbf{s}) \\
      & -\Phi_{1}^{(N)}(z; \mathbf{s})\frac{ \partial  }{ \partial s_{1} } \Phi_{1}^{(N)}(-z; \mathbf{s})+\frac{1}{z}\frac{ \partial  }{ \partial s_{1} } \Phi_{1}^{(N)}(z; \mathbf{s})\frac{ \partial  }{ \partial s_{1} } \Phi_{1}^{(N)}(-z; \mathbf{s}).
    \end{aligned}
  \end{equation}
  Recall that the wave-function
  \[
    w(\mathbf{s}; z):=\exp\left(\sum_{k=0}^{\infty}s_{2k+1}z^{2k+1}\right) \frac{\tau_{BGW}^{(N)}(\mathbf{s}-[z^{-1}]_{\text{odd}})}{\tau_{BGW}^{(N)}(\mathbf{s})}=\exp\left(\sum_{k=0}^{\infty}s_{2k+1}z^{2k+1}\right)\Phi_{1}^{(N)}(-z; \mathbf{s})
  \]
  associated with the tau-function of the KdV hierarchy \(\tau ^{(N)}_{BGW}(\mathbf{s})\) satisfies (see, for example, \cite[\S1.3]{DYZ21})
  \begin{equation}\label{eq:eigen-wave}
    \left( \frac{ \partial ^{2}  }{ \partial s_{1}^{2} }+u \right) w(\mathbf{s}; z) = z^{2}w(\mathbf{s}; z),\qquad u=2 \frac{ \partial ^{2} }{ \partial s_{1}^{2} } \log \tau_{BGW}^{(N)}(\mathbf{s}).
  \end{equation}
  Then we have
  \begin{equation}
    \left( \frac{ \partial ^{2} }{ \partial s_{1}^{2} } +2z\frac{ \partial  }{ \partial s_{1} }+u \right) \Phi_{1}^{(N)}(-z; \mathbf{s})=0,
  \end{equation}
  and by a direct computation,
  \begin{align*}
     \frac{ \partial  }{ \partial s_{1} } \Phi_{1}^{(N)}(z; \mathbf{s})\frac{ \partial  }{ \partial s_{1} } &\Phi_{1}^{(N)}(-z; \mathbf{s}) 
    =  \frac{ \partial  }{ \partial s_{1} } \left( \Phi_{1}^{(N)}(z; \mathbf{s})\frac{ \partial  }{ \partial s_{1} }\Phi_{1}^{(N)}(-z; \mathbf{s}) \right) - \Phi_{1}^{(N)}(z; \mathbf{s})\frac{ \partial ^{2}  }{ \partial s_{1}^{2} }\Phi_{1}^{(N)}(-z; \mathbf{s}) \\
    = & \frac{ \partial  }{ \partial s_{1} } \left( \Phi_{1}^{(N)}(z; \mathbf{s})\frac{ \partial  }{ \partial s_{1} }\Phi_{1}^{(N)}(-z; \mathbf{s}) \right)+\Phi_{1}^{(N)}(z; \mathbf{s})\left(2z\frac{ \partial  }{ \partial s_{1} } +u\right)\Phi_{1}^{(N)}(-z; \mathbf{s}).
  \end{align*}
  Similarly, we have
\begin{align*}
 & \frac{ \partial  }{ \partial s_{1} } \Phi_{1}^{(N)}(-z; \mathbf{s})\frac{ \partial  }{ \partial s_{1} } \Phi_{1}^{(N)}(z; \mathbf{s}) \\
 = & \frac{ \partial  }{ \partial s_{1} } \left( \Phi_{1}^{(N)}(-z; \mathbf{s})\frac{ \partial  }{ \partial s_{1} }\Phi_{1}^{(N)}(z; \mathbf{s}) \right)+\Phi_{1}^{(N)}(-z; \mathbf{s})\left(u-2z\frac{ \partial  }{ \partial s_{1} } \right)\Phi_{1}^{(N)}(z; \mathbf{s}).
\end{align*}
  Averaging the above two formulas, we obtain 
  \begin{equation}\label{eq:Phi22-2}
    \begin{aligned}
      \frac{ \partial  }{ \partial s_{1} } \Phi_{1}^{(N)}(z; \mathbf{s})\frac{ \partial  }{ \partial s_{1} } &\Phi_{1}^{(N)}(-z; \mathbf{s})
      =  \frac{z}{2}\frac{ \partial ^{2} }{ \partial s_{1}^{2} }\Phi_{1,1}^{(N)}(z; \mathbf{s})+zu \Phi_{1,1}^{(N)}(z; \mathbf{s}) \\
      & +z\left( \Phi_{1}^{(N)}(z; \mathbf{s})\frac{ \partial }{ \partial s_{1} }\Phi_{1}^{(N)}(-z; \mathbf{s})-\Phi_{1}^{(N)}(-z; \mathbf{s})\frac{ \partial  }{ \partial s_{1} } \Phi_{1}^{(N)}(z; \mathbf{s}) \right).
    \end{aligned}
  \end{equation}
Combining \eqref{eq:Phi22-1} and \eqref{eq:Phi22-2} gives
  \begin{align*}
    \Phi_{2,2}^{(N)}(z; \mathbf{s}) & =\left(\frac{1}{2}\frac{ \partial ^{2} }{ \partial s_{1}^{2} } +\frac{1}{2}\frac{ \partial  }{ \partial s_{1} } -z^{2}+\frac{1}{4}+u\right)\Phi_{1,1}^{(N)}(z; \mathbf{s}) \\
    & =-z+\left( \frac{1}{4}+u-B^{(N)}(0^{2}; \mathbf{s}) \right) \frac{1}{z} +\sum_{k=1}^{\infty}\left(\frac{1}{2}B^{(N)}(0^{3},k-1;\mathbf{s})\right.\\
    & \left.+\frac{1}{2}B^{(N)}(0^{2},k-1; \mathbf{s})+\left(\frac{1}{4}+u\right)B^{(N)}(0,k-1; \mathbf{s})-B^{(N)}(0,k; \mathbf{s})\right)z^{-2k-1}.
  \end{align*}

\end{proof}

\subsection{Concavity of the generalized BGW correlators}\label{sec:positive-decomposition-concavity}

In this subsection, we prove the strict concavity of the normalized generalized BGW correlators.
To this end, we prove that second differences \(K_{i,j}^{(N)}(\mathbf{d})\) of the generalized BGW coefficients are negative as we have promised at the end of the subsection \ref{sec:generalized-bgw-coefficients}.
\begin{theorem}[=Theorem~\ref{thm:intro-concavity}]\label{thm:convexK}
  Let \(s \in \mathbb{N}_{+}\) and \(0\leq h\leq s+1\). Then the normalized generalized BGW correlators \(C_{h}(\mathbf{d})\) are strictly concave on the set \(\mathcal{P}_{s}\) of partitions of size \(s\).
\end{theorem}
\begin{proof}
  Fix \(0\leq h\leq s+1\). Let \(\mathbf{d}=(d_{1},\dots,d_{l(\mathbf{d})})\) be a partition of size \(s\), and let \(i,j\in[l(\mathbf{d})]\) be distinct indices with \(d_{i},d_{j} \in \mathbb{N}_{+}\). The partitions \(\mathbf{d}\) and \(\mathbf{d}\pm(e_{i}-e_{j})\) have the same size and length, and hence the same positive normalization factor \(\kappa_{\mathbf{d}}= \frac{(2|\mathbf{d}|+l(\mathbf{d})-1)!}{2^{2|\mathbf{d}|+1}}\). Therefore, by \eqref{eq:def-C}, 
  \begin{align*}
    & C_{h}(\mathbf{d}+e_{i}-e_{j})+C_{h}(\mathbf{d}+e_{j}-e_{i})-2C_{h}(\mathbf{d})\\
    = & \kappa_{\mathbf{d}}^{-1}\big(B_{h}(\mathbf{d}+e_{i}-e_{j})+B_{h}(\mathbf{d}+e_{j}-e_{i})-2B_{h}(\mathbf{d})\big)
    = \kappa_{\mathbf{d}}^{-1}K_{i,j;h}(\mathbf{d}).
  \end{align*}
  Thus, it suffices to prove that \(K_{i,j;h}(\mathbf{d})<0\).

  We first derive an expression for \(K_{i,j}^{(N)}(\mathbf{d})\).
  By using Lemma \ref{lem:KN as xi} and equation \eqref{eq:Phi12-2}, we have
  \begin{align} \label{eq:2point-nocontribution}
  	\begin{split}
    &\left( \frac{z_{1}}{z_{2}}-\frac{z_{2}}{z_{1}} \right) ^{2}\xi(z_{1},-z_{2}; \mathbf{s})\xi(z_{2},-z_{1}; \mathbf{s})\\
    =&\frac{1}{8}\left( \Phi_{1,1}^{(N)}(z_{2}; \mathbf{s})\Phi_{2,2}^{(N)}(z_{1}; \mathbf{s})+\Phi_{1,1}^{(N)}(z_{1}; \mathbf{s})\Phi_{2,2}^{(N)}(z_{2}; \mathbf{s})-2\tilde{\Phi}_{1,2}^{(N)}(z_{1}; \mathbf{s})\tilde{\Phi}_{1,2}^{(N)}(z_{2}; \mathbf{s}) \right)
    -\frac{1}{4}\\
    &+\frac{z_1^2+z_2^2}{16z_1z_2}\Big(\Phi_{1,1}^{(N)}(z_{2}; \mathbf{s})\Phi_{2,2}^{(N)}(z_{1}; \mathbf{s})
     +\Phi_{1,1}^{(N)}(z_{1}; \mathbf{s})\Phi_{2,2}^{(N)}(z_{2}; \mathbf{s})-2-2\tilde{\Phi}_{1,2}^{(N)}(z_{1}; \mathbf{s})\tilde{\Phi}_{1,2}^{(N)}(z_{2}; \mathbf{s})\Big),
    \end{split}
  \end{align}
  where $\tilde{\Phi}_{i,j}^{(N)}(z; \mathbf{s})=\Phi_{i,j}^{(N)}(z; \mathbf{s})+1, i,j\in\{ 1,2 \}$.
  The functions \(\tilde{\Phi}_{1,2}^{(N)}(z; \mathbf{s})\), \(\Phi_{1,1}^{(N)}(z; \mathbf{s})\), and \(\Phi_{2,2}^{(N)}(z; \mathbf{s})\) are all odd in \(z\). 
  It follows that the term $-1/4$ and the term in the last line of \eqref{eq:2point-nocontribution} do not contribute to the coefficient of \(z_{1}^{-2x-1}z_{2}^{-2y-1}\) for \(x,y \geq0\). Hence, by Lemma \ref{lem:KN as xi}, we have
  \begin{align}\label{eq:KN-Phi}
  	\begin{split}
    K^{(N)}(d_{i},d_{j}; &\mathbf{s})=-[z_{1}^{-2d_{i}-1}z_{2}^{-2d_{j}-1}]\\ 
    &\big(\Phi_{1,1}^{(N)}(z_{2}; \mathbf{s})\Phi_{2,2}^{(N)}(z_{1}; \mathbf{s})+\Phi_{1,1}^{(N)}(z_{1}; \mathbf{s})\Phi_{2,2}^{(N)}(z_{2}; \mathbf{s})-2\tilde{\Phi}_{1,2}^{(N)}(z_{1}; \mathbf{s})\tilde{\Phi}_{1,2}^{(N)}(z_{2}; \mathbf{s})\big).
    \end{split} 
  \end{align}
  For any partition \(\mathbf{a}=(a_{1},\dots,a_{n})\) and \(k\in \mathbb{N}\), write
  \begin{align}\label{eqn:PhiNpqa}
  	\begin{split}
    \Phi_{p,q,\mathbf{a}}^{(N)}(k) & =\left[ z^{-2k-1} \right] \partial_{\mathbf{a}}\Phi_{p,q}^{(N)}(z;\mathbf{s}) \Big|_{\mathbf{s}=0}, \\
    \tilde{\Phi}_{p,q,\mathbf{a}}^{(N)}(k) & =\left[ z^{-2k-1} \right] \partial_{\mathbf{a}}\tilde{\Phi}_{p,q}^{(N)}(z;\mathbf{s}) \Big|_{\mathbf{s}=0}.
    \end{split}
  \end{align}
  Apply \(\partial_{\mathbf{d}_{[l(\mathbf{d})]\setminus \{ i,j \}}}\) to \eqref{eq:KN-Phi} and set \(\mathbf{s}=0\). Equation \eqref{eq:K-ShiftedK} and the Leibniz rule give 
  \begin{equation}\label{eq:K-Phi}
  	\begin{split}
    K_{i,j}^{(N)}(\mathbf{d})=-\sum_{I\sqcup J=[l(\mathbf{d})]\setminus \{ i,j \}}\Big( &\Phi_{1,1,\mathbf{d}_{I}}^{(N)}(d_{i})\Phi_{2,2,\mathbf{d}_{J}}^{(N)}(d_{j})\\
    &+\Phi_{1,1,\mathbf{d}_{J}}^{(N)}(d_{j})\Phi_{2,2,\mathbf{d}_{I}}^{(N)}(d_{i})-2\tilde{\Phi}_{1,2,\mathbf{d}_{I}}^{(N)}(d_{i})\tilde{\Phi}_{1,2,\mathbf{d}_{J}}^{(N)}(d_{j}) \Big).
    \end{split}
  \end{equation}

  Below,
  we compute and rewrite the right hand side of equation \eqref{eq:K-Phi},
  and then prove each summand in \(K_{i,j}^{(N)}(\mathbf{d})\) is a polynomial in \(-N^{2}\) with negative coefficients.
  Let \(\mathbf{a}\) be a partition and \(k\geq 1\). Taking $m=0$ in \eqref{eq:Virasoro-B} gives $B^{(N)}(0,\mathbf{d})=\chi(\mathbf{d})B^{(N)}(\mathbf{d})$.
  By Lemma~\ref{lem:Phis-Bs} and the notation \eqref{eqn:PhiNpqa}, 
  terms appearing on the right-hand side of equation \eqref{eq:K-Phi} take the following form
  \begin{align}
    \Phi_{1,1,\mathbf{a}}^{(N)}(k) & =B^{(N)}(\mathbf{a},0,k-1) \label{eq:Phi11-B}, \\
    \tilde{\Phi}_{1,2,\mathbf{a}}^{(N)}(k) & = -\frac{1}{2}(2k+1+\chi(\mathbf{a}))B^{(N)}(\mathbf{a},0,k-1) \label{eq:Phi12-B}, \\
    \begin{split}
    \Phi_{2,2,\mathbf{a}}^{(N)}(k) & = \frac{2(2k+1+\chi(\mathbf{a}))^{2}-1}{4}B^{(N)}(\mathbf{a},0,k-1)-B^{(N)}(\mathbf{a},0,k) \\
    &\ \ \ \  +2\sum_{I\sqcup J=[l(\mathbf{a})]}B^{(N)}(\mathbf{a}_{I},0^{2})B^{(N)}(\mathbf{a}_{J},0,k-1). \label{eq:Phi22-B-1}
    \end{split}
  \end{align}
  We further deal with the $\Phi_{2,2,\mathbf{a}}^{(N)}(k)$.
  Applying \eqref{eq:Phi12-2}, one has
  \begin{equation}\label{eq:Phi1122}
  	\begin{split}
    \Phi_{1,1}^{(N)}(z; \mathbf{s})\Phi_{2,2}^{(N)}(z; \mathbf{s}) &=-\Phi_{1,2}^{(N)}(z; \mathbf{s})\Phi_{1,2}^{(N)}(-z; \mathbf{s})\\
    & =\Phi_{1,2}^{(N)}(z; \mathbf{s})(\Phi_{1,2}^{(N)}(z; \mathbf{s})+2)=\tilde{\Phi}_{1,2}^{(N)}(z; \mathbf{s})^{2}-1.
    \end{split}
  \end{equation}
In the proof of Lemma~\ref{lem:Phis-Bs}, we derive expressions for \(\Phi_{1,2}^{(N)}(z; \mathbf{s})\) and \(\Phi_{2,2}^{(N)}(z; \mathbf{s})\) in terms of \(\Phi_{1,1}^{(N)}(z; \mathbf{s})\). Consequently, the identity \eqref{eq:Phi1122} can be rewritten entirely in terms of \(\Phi_{1,1}^{(N)}(z; \mathbf{s})\),
which yields
  \begin{equation}
    \frac{1}{2}\Phi_{1,1}^{(N)}(z; \mathbf{s})\frac{ \partial ^{2} }{ \partial s_{1}^{2} } \Phi_{1,1}^{(N)}(z; \mathbf{s})-\frac{1}{4}\left(\frac{ \partial  }{ \partial s_{1} } \Phi_{1,1}^{(N)}(z; \mathbf{s})\right)^{2}+(u-z^{2})\Phi_{1,1}^{(N)}(z; \mathbf{s})^{2}+1=0.
  \end{equation}
  Differentiating with respect to \(s_{1}\) further gives the following identity
  \begin{align*}
    0=\Phi_{1,1}^{(N)}(z; \mathbf{s})\left(\frac{1}{2}\frac{ \partial ^{3} }{ \partial s_{1}^{3} } \Phi_{1,1}^{(N)}(z; \mathbf{s})+2(u-z^{2})\frac{ \partial  }{ \partial s_{1} } \Phi_{1,1}^{(N)}(z; \mathbf{s})+\Phi_{1,1}^{(N)}(z; \mathbf{s})\frac{ \partial u }{ \partial s_{1} } \right).
  \end{align*}
  Canceling the factor \(\Phi_{1,1}^{(N)}(z; \mathbf{s})\) and comparing the coefficients of \(z^{-2k-1}\) in the above equation gives the following relation
  \begin{align*}
    B^{(N)}(0^{2},k; \mathbf{s}) & =\frac{1}{4}B^{(N)}(0^{4},k-1; \mathbf{s})+2B^{(N)}(0^{2}; \mathbf{s})B^{(N)}(0^{2},k-1; \mathbf{s}) \\
    &\quad +B^{(N)}(0^{3}; \mathbf{s})B^{(N)}(0,k-1; \mathbf{s}).
  \end{align*}
  Applying \(\partial_{\mathbf{a}}\) to both sides of the above equation and then setting \(\mathbf{s}=0\), we obtain 
  \begin{align*}
    B^{(N)}(\mathbf{a},0^{2},k)  =&\frac{1}{4}B^{(N)}(\mathbf{a},0^{4},k-1)+ \sum_{I\sqcup J=[l(\mathbf{a})]}
    \big(2B^{(N)}(\mathbf{a}_{I},0^{2})B^{(N)}(\mathbf{a}_{J},0^{2},k-1) \\
    & +B^{(N)}(\mathbf{a}_{I},0^{3})B^{(N)}(\mathbf{a}_{J},0,k-1)\big).
  \end{align*}
  Using the string equation $B^{(N)}(0,\mathbf{d})=\chi(\mathbf{d}) \cdot B^{(N)}(\mathbf{d})$ to eliminate one zero insertion, we rewrite this identity as
  \begin{align*}
    B^{(N)}(\mathbf{a},0,k)  =&\frac{(2k+1+\chi(\mathbf{a}))(2k+\chi(\mathbf{a}))}{4}B^{(N)}(\mathbf{a},0,k-1) \\
    & +\sum_{I\sqcup J=[l(\mathbf{a})]}\left(2-\frac{2+\chi(\mathbf{a}_{I})}{2k+2+\chi(\mathbf{a})}\right)B^{(N)}(\mathbf{a}_{I},0^{2})B^{(N)}(\mathbf{a}_{J},0,k-1).
  \end{align*}
  Substituting this result into \eqref{eq:Phi22-B-1} yields 
  \begin{equation}\label{eq:Phi22-B}
    \begin{aligned}
      \Phi_{2,2,\mathbf{a}}^{(N)}(k) =& \frac{(2k+1+\chi(\mathbf{a}))(2k+2+\chi(\mathbf{a}))-1}{4}B^{(N)}(\mathbf{a},0,k-1) \\
      & +\sum_{I\sqcup J=[l(\mathbf{a})]}\frac{2+\chi(\mathbf{a}_{I})}{2k+2+\chi(\mathbf{a})}B^{(N)}(\mathbf{a}_{I},0^{2})B^{(N)}(\mathbf{a}_{J},0,k-1).
    \end{aligned}
  \end{equation}
  
  By combining \eqref{eq:Phi11-B}, \eqref{eq:Phi12-B}, and \eqref{eq:Phi22-B}, we obtain, for partitions \(\mathbf{a},\mathbf{b}\) and positive integers \(k,l\),
  \begin{align}
    & \Phi_{1,1,\mathbf{a}}^{(N)}(k)\Phi_{2,2,\mathbf{b}}^{(N)}(l)+\Phi_{1,1,\mathbf{b}}^{(N)}(l)\Phi_{2,2,\mathbf{a}}^{(N)}(k)-2\tilde{\Phi}_{1,2,\mathbf{a}}^{(N)}(k)\tilde{\Phi}_{1,2,\mathbf{b}}^{(N)}(l) \label{eq:summand-Phi} \\
    = & \frac{(\chi(k,\mathbf{a})-\chi(l,\mathbf{b}))^{2}+\chi(k,\mathbf{a})+\chi(l,\mathbf{b})-2}{4}B^{(N)}(\mathbf{a},0,k-1)B^{(N)}(\mathbf{b},0,l-1)\label{eq:Phi-quadratic-C}  \\
    &+ B^{(N)}(\mathbf{a},0,k-1)\beta_{\mathbf{b}}^{(N)}(l)+B^{(N)}(\mathbf{b},0,l-1)\beta_{\mathbf{a}}^{(N)}(k),\label{eq:Phi-cubic-C}
  \end{align}
  where the function $\beta_{\mathbf{a}}^{(N)}(k)$ is
  \begin{equation}\label{eq:def-beta}
    \beta_{\mathbf{a}}^{(N)}(k)
    =\sum_{I\sqcup J=[l(\mathbf{a})]}\frac{2+\chi(\mathbf{a}_{I})}{2k+2+\chi(\mathbf{a})}B^{(N)}(\mathbf{a}_{I},0^{2})B^{(N)}(\mathbf{a}_{J},0,k-1).
  \end{equation}
  
  Recall the formula \eqref{eq:K-Phi}.
  The negative second difference $-K_{i,j}^{(N)}(\mathbf{d})$ is equal to
  \begin{equation}\label{eqn:rhs of K} 
  	\begin{split}
  		\sum_{I\sqcup J=[l(\mathbf{d})]\setminus \{ i,j \}}\Big( \Phi_{1,1,\mathbf{d}_{I}}^{(N)}(d_{i})\Phi_{2,2,\mathbf{d}_{J}}^{(N)}(d_{j})
  		+\Phi_{1,1,\mathbf{d}_{J}}^{(N)}(d_{j})\Phi_{2,2,\mathbf{d}_{I}}^{(N)}(d_{i})-2\tilde{\Phi}_{1,2,\mathbf{d}_{I}}^{(N)}(d_{i})\tilde{\Phi}_{1,2,\mathbf{d}_{J}}^{(N)}(d_{j}) \Big).
  	\end{split}
  \end{equation}
  Let \(I,J\) be subsets of \([l(\mathbf{d})]\) satisfying \(I\sqcup J=[l(\mathbf{d})]\setminus \{ i,j \}\), and  \(\mathbf{a}=\mathbf{d}_{I},\mathbf{b}=\mathbf{d}_{J},k=d_{i},l=d_{j}\). 
  Then each term appearing in the formula \eqref{eqn:rhs of K} is of the form \eqref{eq:summand-Phi}.
  Since \(l,k\geq1\), we have \(\chi(k,\mathbf{a}),\chi(l,\mathbf{b})\geq3\). By Proposition~\ref{prp:Bhd-positive},
  a term as \eqref{eq:Phi-quadratic-C} contributes a polynomial of degree \(|\mathbf{d}|\) in \(-N^{2}\) and the coefficients are positive in degrees from $0$ to $|\mathbf{d}|$,
  whereas a term as \eqref{eq:Phi-cubic-C} contributes a polynomial of degree \(|\mathbf{d}|+1\) in \(-N^{2}\) and the coefficients are positive in degrees from $0$ to $|\mathbf{d}|+1$.
  Consequently, \(-K_{i,j}^{(N)}(\mathbf{d})\) is a polynomial in $-N^2$ with positive coefficients in every degree from $0$ to $|\mathbf{d}|+1$; that is, in the notation of \eqref{def:KijN as Kh},
  we have $K_{i,j;h}(\mathbf{d})<0$ with $0\leq h \leq |\mathbf{d}|+1$.
\end{proof}

\section{Strict Schur-concavity and bounds for the generalized BGW correlators}\label{sec:schur-concavity}
In this section, we apply Theorem~\ref{thm:convexK} to establish the strict Schur-concavity of the normalized generalized BGW correlators \(C_{h}\) and derive bounds for them.

We begin by introducing the notion of Schur-concavity in the setting of partitions, using the dominance order given in Definition~\ref{def:order-Macdonald}. For a treatment of Schur-concavity in a more general setting, we refer the reader to \cite{MOA11}.

\begin{definition}\label{def:Schur-concave}
  Let \(s\geq1\). A function \(f:\mathcal{P}_{s}^{+}\to \mathbb{R}\) is called Schur-concave if \(\mathbf{a}\trianglerighteq\mathbf{b}\) implies \(f(\mathbf{a})\leq f(\mathbf{b})\) for all \(\mathbf{a},\mathbf{b} \in \mathcal{P}_{s}^{+}\). It is called strictly Schur-concave if, in addition, \(\mathbf{a}\triangleright \mathbf{b}\) implies \(f(\mathbf{a})<f(\mathbf{b})\). A function \(f\) is called (strictly) Schur-convex if \(-f\) is (strictly) Schur-concave.
\end{definition}

For each integer \(l\geq2\) and distinct indices \(i,j\in[l]\), define \(R_{i,j}:\mathbb{Z}^{l}\to \mathbb{Z}^{l}\) by
\begin{equation}
  R_{i,j}(\mathbf{a})=\mathbf{a}+e_{i}-e_{j}.
\end{equation}

\begin{definition}\label{def:transfer}
  Let \(\mathbf{d}=(d_{1},\dots,d_{l})\) be a partition in \(\mathcal{P}_{s}\). For indices \(i,j\in[l]\) satisfying \(d_{j}\geq d_{i}+2\), the operation
  \begin{equation}
    \mathbf{d}\mapsto R_{i,j}(\mathbf{d})
  \end{equation}
  is called a transfer from \(\mathbf{d}\) to \(R_{i,j}(\mathbf{d})\). 
\end{definition}

If \(\mathbf{a}\) is obtained from \(\mathbf{b}\) by a single transfer, then \(\mathbf{a}\triangleleft\mathbf{b}\). Conversely, we have the following lemma; see also \cite[Lemma~5.D.1]{MOA11} and \cite[(1.15)]{Mac98}.
\begin{lemma}\label{lem:muirhead}
  Let \(\mathbf{a},\mathbf{b} \in \mathcal{P}_{s,l}\) satisfy \(\mathbf{a}\triangleleft \mathbf{b}\). Then \(\mathbf{a}\) can be derived from \(\mathbf{b}\) by a finite sequence of transfers.
\end{lemma}
\begin{proof}
  Starting with \(\mathbf{c}=\mathbf{b}^{\downarrow}\), the transfer at each step is chosen to preserve
  \begin{equation}
    \sum_{i=1}^{k}c_{i}^{\downarrow}\geq\sum_{i=1}^{k}a_{i}^{\downarrow}.
  \end{equation}
  At each step, write \(\mathbf{c}\) and \(\mathbf{a}\) in weakly decreasing order. If \(\mathbf{c}^{\downarrow}\neq \mathbf{a}^{\downarrow}\), let \(j\) be the first index at which they differ. Dominance gives \(c_{j}^{\downarrow}>a_{j}^{\downarrow}\), and equality of the total sums ensures that there exists \(i>j\) such that \(c_{i}^{\downarrow}<a_{i}^{\downarrow}\). We choose \(i\) to be the smallest such index. Since \(c_{i}^{\downarrow}<a_{i}^{\downarrow}\leq a_{j}^{\downarrow}<c_{j}^{\downarrow}\), the operation
  \[
    \mathbf{c}\mapsto R_{i,j}(\mathbf{c}^{\downarrow})
  \]
  is a transfer. By construction, the resulting partition still dominates \(\mathbf{a}\), that is, \(\mathbf{a}\trianglelefteq R_{i,j}(\mathbf{c}^{\downarrow})\). Each transfer decreases
  \[
    \sum_{v=1}^{l(\mathbf{a})}|c_{v}^{\downarrow}-a_{v}^{\downarrow}|
  \]
  by at least two after rearrangement. This nonnegative integer therefore reaches zero after finitely many steps, at which point \(\mathbf{c}=\mathbf{a}\).
\end{proof}

\begin{theorem}[=Theorem~\ref{thm:intro-bound}]\label{thm:monotone-pseudo-adjacent}
  Let \(s \in \mathbb{N}_{+}\). The function \(C_{h}:\mathcal{P}_{s}^{+}\to \mathbb{R}\) is strictly Schur-concave for \(h \in \{ 0,\dots,s+1 \}\). Consequently, we have the following bounds for the generalized BGW correlators:
  \begin{equation}\label{eqn:bounds Ch}
    C_{h}(s)\leq C_{h}(\mathbf{d})\leq C_{h}(1^{s}).
  \end{equation}
\end{theorem}
\begin{proof}
  Let \(\mathbf{d},\mathbf{d}'\in \mathcal{P}_{s}^{+}\) and \(\mathbf{d}\triangleright \mathbf{d}'\). 
  We consider their zero-padded partitions
  \[
    \rho=(0^{s-l(\mathbf{d})},\mathbf{d}),\qquad \rho'=(0^{s-l(\mathbf{d}')},\mathbf{d}')
  \]
  in \(\mathcal{P}_{s,s}\), where \(\rho\triangleright \rho'\) by Remark~\ref{rem:order-iso}.
  Since the normalization constants $\kappa_{\mathbf{d}}= \frac{(2|\mathbf{d}|+l(\mathbf{d})-1)!}{2^{2|\mathbf{d}|+1}}$ for \(\rho\) and \(\rho'\) are identical:
  \begin{align}\label{eqn:kappa same}
  	\kappa_{\rho}=\kappa_{\rho'}= \frac{(3s-1)!}{2^{2s+1}},
  \end{align}
  to prove $C_h(\rho)<C_h(\rho')$,
  it suffices to prove $B_h(\rho)<B_h(\rho')$.
  
      By Lemma~\ref{lem:muirhead}, there is a finite sequence of transfers from \(\rho\) to \(\rho'\). It therefore suffices to consider the case \(\rho'=R_{i,j}(\rho)\) for some transfer \(R_{i,j}\) and prove that \(B_{h}(\rho')>B_{h}(\rho)\). By the symmetry of the second difference of the generalized BGW coefficients \(K_{i,j}^{(N)}(\mathbf{d})\) under interchange of the \(i\)-th and \(j\)-th entries, we have
  \begin{equation}\label{eq:B-K}
    B_{h}(\rho')-B_{h}(\rho)=-\frac{1}{2}\sum_{k=0}^{\rho_{j}'-\rho_{i}'}K_{i,j;h}\big(\rho'+k(e_{i}-e_{j})\big).
  \end{equation}
  By Theorem~\ref{thm:convexK},
  each second difference in the above sum is strictly negative.
  Consequently, we have
  \[
    B_{h}(\rho')>B_{h}(\rho).
  \]
  By the string equation $C_h(\mathbf{d})=C_h(0,\mathbf{d})$ and \eqref{eqn:kappa same}, the Schur-concavity of \(C_{h}\) is given by
  \[
    C_{h}(\mathbf{d})=C_{h}(\rho)=\frac{B_{h}(\rho)}{\kappa_{\rho}}< \frac{B_{h}(\rho')}{\kappa_{\rho'}}=C_{h}(\rho')=C_{h}(\mathbf{d}').
  \]
  Since \(|\mathbf{d}|\trianglerighteq \mathbf{d}\trianglerighteq 1^{|\mathbf{d}|}\), the Schur-concavity of $C_h$ proves the bounds \eqref{eqn:bounds Ch} for the generalized BGW correlators.
\end{proof}

Setting \(h=0\) and writing \(C(\mathbf{d})=C_{0}(\mathbf{d})\), we prove the conjectured upper bound for the normalized BGW numbers proposed by Guo--Norbury--Yang--Zagier \cite{GNYZ25}. 

\begin{corollary}[Conjecture~1 in \cite{GNYZ25}]
	\label{cor:Bound-BGW}
  For every integer \(g\geq2\) and every positive partition \(\mathbf{d} \) of \(g-1\), we have
  \begin{equation}
    C(g-1)\leq C(\mathbf{d})\leq C(1^{g-1}).
  \end{equation}
\end{corollary}

\section{Strict monotonicity of ordinary BGW correlators}
\label{sec:strict-monotonicity}
In this section, we set \(h=0\) and focus on the ordinary BGW model. We suppress the subscript \(h=0\) for convenience.
Following the notations in Guo--Norbury--Yang--Zagier \cite{GNYZ25}, we write
\[
  B(\mathbf{d})=B_{0}(\mathbf{d}),\qquad C(\mathbf{d})=C_{0}(\mathbf{d})
\]
and call these quantities the BGW coefficients and the normalized BGW correlators, respectively.  To state the monotonicity result, we first recall the total order on \(\mathcal{P}_{s}^{+}\) introduced by Guo--Norbury--Yang--Zagier \cite{GNYZ25}.
\begin{definition}\label{def:order-GNYZ}
  For
  \[
    \mathbf{d}=(d_{1},\dots,d_{l}),\mathbf{d}'=(d_{1}',\dots,d_{l'}')\in \mathcal{P}_{s}^{+},
  \]
  we write \(\mathbf{d}\prec\mathbf{d}'\) if either \(l<l'\), or \(l=l'\) and \(d_{j}^{\uparrow}<d_{j}'^{\uparrow}\) at the first index \(j\) for which \(d_{j}^{\uparrow}\neq d_{j}'^{\uparrow}\). 
\end{definition}
Note that this order is the reverse of the reverse colexicographic order defined in Definition~\ref{def:order-Macdonald}. We therefore call it the colexicographic order. The following theorem proves the strict monotonicity conjecture of Guo--Norbury--Yang--Zagier \cite{GNYZ25}.
\begin{theorem}[=Theorem~\ref{thm:intro-monotonicity}]\label{thm:monotonicity-BGW}
  Let \(g\geq2\), and let \(\mathbf{d}\prec \mathbf{d}'\) be positive partitions of \(g-1\). Then \(C(\mathbf{d})<C(\mathbf{d}')\).
\end{theorem}

\subsection{Adjacent partitions}\label{sec:adjacent-partition}
In this subsection, we characterize adjacent partitions in the colexicographic order and the maximum positive partition of a given size and length.
\begin{proposition}\label{prop:adjacent-partitions}
  Let \(\mathbf{d}=(d_{1}\leq\dots\leq d_{l})\) be a positive partition of \(g-1\). Let \(m(\mathbf{d})\) be the largest index \(j\in[l]\) such that 
  \[
    (l-j+1)(d_{j}+1)\leq d_{j}+\cdots+d_{l},
  \]
  and set \(m(\mathbf{d})=0\) if no such index exists.

  \begin{enumerate}[(1)]
    \item If \(m(\mathbf{d})>0\), the adjacent partition \(\mathbf{d}'\) of \(\mathbf{d}\) such that \(\mathbf{d}'\succ \mathbf{d}\) is
    \[
      \mathbf{d}'=(d_{1},\dots,d_{m(\mathbf{d})-1},(d_{m(\mathbf{d})}+1)^{l-m(\mathbf{d})},d'),
    \]
    where \(d'=g-1-d_{1}-\cdots-d_{m(\mathbf{d})-1}-(l-m(\mathbf{d}))(d_{m(\mathbf{d})}+1)\).
    \item If \(m(\mathbf{d})=0\) and \(l< g-1\), then \(\mathbf{d}\) is the maximum element in \(\mathcal{P}_{g-1,l}^{+}\), and the adjacent partition \(\mathbf{d}'\) of \(\mathbf{d}\) such that \(\mathbf{d}'\succ \mathbf{d}\) is the minimum element in \(\mathcal{P}_{g-1,l+1}^{+}\), namely \((1^{l},g-1-l)\). Furthermore, if we write \(g-1=d_{1}l+t\) with \(0\leq t<l\), then we have the explicit expression for \(\mathbf{d}\):
    \begin{equation}
      \mathbf{d}=(d_{1}^{l-t},(d_{1}+1)^{t}).
    \end{equation}
  \end{enumerate}
\end{proposition}
\begin{proof}
  Note that a pair of partitions \(\mathbf{d}\prec \mathbf{d}'\) is adjacent if and only if for every partition \(\mathbf{e}\) such that \(\mathbf{d}\preceq\mathbf{e}\preceq\mathbf{d}'\), one has \(\mathbf{e}=\mathbf{d}\) or \(\mathbf{e}=\mathbf{d}'\). 
  
  (1). First assume that \(m=m(\mathbf{d})>0\). Any partition \(\mathbf{e}\) satisfying \(\mathbf{d}\preceq\mathbf{e}\preceq\mathbf{d}'\) has the following form
    \[
      \mathbf{e}=(d_{1}\leq \cdots\leq d_{m-1}\leq e_{m}\leq\cdots \leq e_{l}).
    \]
    Suppose that \(\mathbf{e}\neq \mathbf{d}\), and set $j=\min\{ k\big|e_{k}>d_{k}\}$.
    We have \(j\geq m\) and \(e_{k}=d_{k}\) for \(k<j\). If \(j>m\), then the maximality of \(m(\mathbf{d})\) gives
    \begin{equation}\label{eq:5.3-contra}
    \begin{aligned}
      e_{l} & = g-1-d_{1}-\cdots-d_{j-1}-e_{j}-\cdots-e_{l-1} \\
      & =d_{j}+\cdots+d_{l}-e_{j}-\cdots-e_{l-1} \\
      & <  (l-j+1)(d_{j}+1)-(l-j)e_{j}
      \leq d_{j}+1\leq e_{j},
    \end{aligned}
  \end{equation}
    which is impossible. Hence \(j=m\). Since \(e_{k}=d_{k}'\) for all \(k\leq m-1\), we have 
    \[
      d_{m}<e_{m}\leq d_{m}'=d_{m}+1.
    \]
    Hence \(e_{m}=d_{m}+1\). Let \(K\) be an integer such that \(m< K<l\). If \(e_{k}=d_{m}+1\) for all \(m\leq k<K\), then 
    \[
      d_{m}+1=e_{K-1}\leq e_{K}\leq d_{K}'=d_{m}+1.
    \]
    It follows inductively that \(e_{K}=d_{m}+1\) for all \(m\leq K<l\). Hence \(\mathbf{e}=\mathbf{d}'\), proving that \(\mathbf d'\) is the adjacent partition of \(\mathbf{d}\) such that \(\mathbf{d}'\succ\mathbf{d}\).
    
    (2). Suppose that \(m(\mathbf{d})=0\) and \(l<g-1\). Let \(\mathbf{e}=(e_{1},\dots,e_{l})\in \mathcal{P}_{g-1,l}^{+}\) such that \(\mathbf{e}\succ \mathbf{d}\). Then \(j:=\min \{ k\mid e_{k}>d_{k} \}\) exists, and the inequality \eqref{eq:5.3-contra} holds since \(m(\mathbf{d})=0\), which is impossible. Hence \(\mathbf{d}\) is the maximum element in \(\mathcal{P}_{g-1,l}^{+}\), and \(\mathbf{d}'\) is the minimum element of \(\mathcal{P}_{g-1,l+1}^{+}\), namely \((1^{l},g-1-l)\). Assume that \(d_{l}-d_{1}\geq2\) and denote \(k=\max \{ j \in[l]\mid d_{j}=d_{1} \}\).
    Then we have
    \begin{align*}
      d_{k}+\cdots+d_{l} & \geq d_{k}+(l-k-1)(d_{k}+1)+d_{k}+2 \\
      & =(l-k+1)(d_{k}+1).
    \end{align*}
    Hence \(m(\mathbf{d})\geq k>0\), contradicting the assumption that \(m(\mathbf{d})=0\). Therefore, \(d_{l}-d_{1}\leq1\), which forces \(\mathbf{d}=(d_{1}^{l-t},(d_{1}+1)^{t})\).
\end{proof}

We conclude this subsection by introducing notation used in the subsequent proofs. Set
\begin{equation}\label{eq:def-Z}
  Z(k,l,\mathbf{a},\mathbf{b})=\frac{\big(\chi(k,\mathbf{a})-\chi(l,\mathbf{b})\big)^{2}+\chi(k,\mathbf{a})+\chi(l,\mathbf{b})-2}{4},
\end{equation}
where $\chi(\mathbf{d})=2|\mathbf{d}|+l(\mathbf{d})$.
Equations \eqref{eq:K-Phi} and \eqref{eq:Phi-quadratic-C}--\eqref{eq:Phi-cubic-C} give the decomposition
\begin{align}
  K_{i,j}(\mathbf{d})= & -\sum_{I\sqcup J=[l(\mathbf{d})]\setminus \{ i,j \}} Z(d_{i},d_{j},\mathbf{d}_{I},\mathbf{d}_{J})B(\mathbf{d}_{I},0,d_{i}-1)B(\mathbf{d}_{J},0,d_{j}-1) \label{eq:def-K2} \\
  & -\sum_{I\sqcup J=[l(\mathbf{d})]\setminus \{ i,j \}}(B(\mathbf{d}_{I},0,d_{i}-1)\beta_{\mathbf{d}_{J}}(d_{j})+B(\mathbf{d}_{J},0,d_{j}-1)\beta_{\mathbf{d}_{I}}(d_{i})), \label{eq:def-K3}
\end{align}
where
\begin{equation}
	\beta_{\mathbf{a}}(k)
	=\sum_{I\sqcup J=[l(\mathbf{a})]}\frac{2+\chi(\mathbf{a}_{I})}{2k+2+\chi(\mathbf{a})}B(\mathbf{a}_{I},0^{2})B(\mathbf{a}_{J},0,k-1).
\end{equation}
We denote the expressions in \eqref{eq:def-K2} and \eqref{eq:def-K3} by \(K_{i,j}^{[2]}(\mathbf{d})\) and \(K_{i,j}^{[3]}(\mathbf{d})\), respectively.

For \(p=2,3\), expand the above expression \eqref{eq:def-K2}-\eqref{eq:def-K3} for \(K_{i,j}^{[p]}(\mathbf{d})\) in the normalized BGW correlators \(C\), using \(B(\mathbf{a})=\kappa_{\mathbf{a}}C(\mathbf{a})\) and the definition of \(\beta_{\mathbf{a}}(k)\). 
We then use the superscript $\star$ to indicate that the normalized BGW correlators are omitted, so that attention is restricted to the coefficients alone.
Explicitly,
denote the sum of the coefficients in this expansion \eqref{eq:def-K2}-\eqref{eq:def-K3} by \(K_{i,j}^{[p],\star}(\mathbf{d})\),
i.e.,
\begin{align}
  K_{i,j}^{[2],\star}(\mathbf{d}) & :=-\sum_{I\sqcup J=[l(\mathbf{d})]\setminus \{ i,j \}} Z(d_{i},d_{j},\mathbf{d}_{I},\mathbf{d}_{J})\kappa_{(\mathbf{d}_{I},0,d_{i}-1)}\kappa_{(\mathbf{d}_{J},0,d_{j}-1)}, \label{eq:def-K2star} \\
  K_{i,j}^{[3],\star}(\mathbf{d}) & :=-\sum_{I\sqcup J=[l(\mathbf{d})]\setminus \{ i,j \}}
  \big(\kappa_{(\mathbf{d}_{I},0,d_{i}-1)}\beta^{\star}_{\mathbf{d}_{J}}(d_{j})+\kappa_{(\mathbf{d}_{J},0,d_{j}-1)}\beta_{\mathbf{d}_{I}}^{\star}(d_{i})\big), \label{eq:def-K3star}
\end{align}
where
\[
  \beta_{\mathbf{a}}^{\star}(k)=\sum_{I\sqcup J=[l(\mathbf{a})]} \frac{2+\chi(\mathbf{a}_{I})}{2k+2+\chi(\mathbf{a})}\kappa_{(\mathbf{a}_{I},0,0)}\kappa_{(\mathbf{a}_{J},0,k-1)}.
\]
Let \(\mathbf{a},\mathbf{b}\) be partitions of the same size and length with \(\mathbf{a}\triangleright\mathbf{b}\). Fix a transfer chain
\begin{equation}
  \mathbf{a}=\mathbf{a}^{(0)}\prec\cdots\prec\mathbf{a}^{(q)}=\mathbf{b}, \qquad \mathbf{a}^{(j)}=R_{k_{j},m_{j}}(\mathbf{a}^{(j-1)}).
\end{equation}
Applying the formula \eqref{eq:B-K} to each step and summing along the chain gives
\begin{equation}
  B(\mathbf{b})-B(\mathbf{a})=-\frac{1}{2}\sum_{j=1}^{q}\sum_{k=0}^{a^{(j)}_{m_{j}}-a_{k_{j}}^{(j)}}K_{k_{j},m_{j}}\big(\mathbf{a}^{(j)}+k(e_{k_{j}}-e_{m_{j}})\big).
\end{equation}
Using this fixed expansion, we define the quadratic and cubic components of \(B(\mathbf{b})-B(\mathbf{a})\) and their coefficient sums as follows, for \(p=2,3\):
\begin{align}
  (B(\mathbf{b})-B(\mathbf{a}))^{[p]} & =-\frac{1}{2}\sum_{j=1}^{q}\sum_{k=0}^{a^{(j)}_{m_{j}}-a_{k_{j}}^{(j)}}K_{k_{j},m_{j}}^{[p]}(\mathbf{a}^{(j)}+k(e_{k_{j}}-e_{m_{j}})), \label{eq:alpha-j} \\
  (B(\mathbf{b})-B(\mathbf{a}))^{[p],\star} & =-\frac{1}{2}\sum_{j=1}^{q}\sum_{k=0}^{a^{(j)}_{m_{j}}-a_{k_{j}}^{(j)}}K_{k_{j},m_{j}}^{[p],\star}(\mathbf{a}^{(j)}+k(e_{k_{j}}-e_{m_{j}})). \label{eq:alpha-j-star}
\end{align}

\subsection{Strict monotonicity in the equal-length case}\label{sec:case-equal-length}
In this subsection, we prove strict monotonicity of the BGW correlators in the equal-length case.

\begin{theorem}\label{thm:equal-length}
  Let \(g\geq2\), and let \(\mathbf{d}\prec \mathbf{d}'\) be adjacent positive partitions of \(g-1\) of equal length. 
  We have \(C(\mathbf{d})<C(\mathbf{d}')\).
\end{theorem}
\begin{proof}
  If \(\mathbf{d}'\) can be obtained from \(\mathbf{d}\) by a single transfer, the desired inequality follows from Theorem~\ref{thm:monotone-pseudo-adjacent}. We therefore assume that no such transfer exists. 
  
  Let \(\mathbf{d}=(d_{1}\leq\cdots\leq d_{l})\) be a nonempty positive partition and set \(m=m(\mathbf{d}),n=l-m\). Set \(\eta:=(d_{m+1},\dots,d_{l})\). Since \(m(\eta)=0\), by Proposition~\ref{prop:adjacent-partitions}, we know that \(\eta\) is the maximum element in \(\mathcal{P}_{|\eta|,n}^{+}\). Write \(|\eta|=d_{m+1}n+r\) with \(0\leq r<n\). Then we have
  \[
    \eta=(d_{m+1}^{n-r},(d_{m+1}+1)^{r}).
  \]
  For convenience, denote \(\pi=(d_{1},\dots,d_{m-1})\). Then we have
  \[
    \mathbf{d}=(\pi,d_{m},\eta),
  \]
  and \(\mathbf{d}'\) is given by
  \[
    \mathbf{d}'=(\pi,(d_{m}+1)^{n},n(d_{m+1}-d_{m}-1)+d_{m}+r).
  \]
  To compare $\mathbf{d}$ and $\mathbf{d}'$,
  we construct an auxiliary partition
  \begin{equation}
    \mathbf{d}''=(\pi,d_{m},(d_{m}+1)^{n-1},n(d_{m+1}-d_{m}-1)+d_{m}+r+1)
  \end{equation}
  such that $\mathbf{d}'$ can be obtained from $\mathbf{d}''$ by a single transfer in the sense of Definition \ref{def:transfer},
  i.e., $\mathbf{d}'=R_{m,l}(\mathbf{d}'')$.
  Since we have excluded the case where \(\mathbf{d}'\) is obtained from \(\mathbf{d}\) by a single transfer, we have \(\mathbf{d}\neq \mathbf{d}''\). 
  Moreover, one can directly check that \(\mathbf{d}\triangleleft \mathbf{d}''\). 
  Then by Lemma \ref{lem:muirhead},
  there exists a transfer chain from $\mathbf{d}''$ to $\mathbf{d}$.
  We fix it as follows
  \begin{align}
    \mathbf{d}''=\mathbf{d}^{(0)}\to\cdots\to \mathbf{d}^{(q)}=\mathbf{d}, \qquad\qquad \mathbf{d}^{(j)}=R_{k_{j},m_{j}}(\mathbf{d}^{(j-1)}).\label{eq:d-transfer-chain}
  \end{align}
  Denote
  \begin{equation}\label{eq:def-Delta-Delta'}
  	\Delta(\mathbf{d})=B(\mathbf{d})-B(\mathbf{d}''),\qquad\Delta'(\mathbf{d})=B(\mathbf{d}')-B(\mathbf{d}'').
  \end{equation}
  Then using the chain $\mathbf{d}'=R_{m,l}(\mathbf{d}'')$ and the chain in \eqref{eq:d-transfer-chain}, 
  we can define \(\Delta ^{[p]}(\mathbf{d})\), \(\Delta ^{[p],\star}(\mathbf{d})\), \((\Delta')^{[p]}(\mathbf{d})\) and \((\Delta')^{[p],\star}(\mathbf{d})\) as in \eqref{eq:alpha-j} and \eqref{eq:alpha-j-star} for \(p=2,3\).

  Since \(\mathbf{d}\) and \(\mathbf{d}'\) have the same size and length, their normalization factors $\kappa_{\mathbf{d}}, \kappa_{\mathbf{d}'}$ agree. 
  Hence
  \begin{equation}\label{eq:C-B-I}
    C(\mathbf{d}')-C(\mathbf{d})= \frac{B(\mathbf{d}')-B(\mathbf{d})}{\kappa_{\mathbf{d}}}=\frac{\Delta'(\mathbf{d})-\Delta(\mathbf{d})}{\kappa_{\mathbf{d}}}.
  \end{equation}
  Thus to prove the strict monotonicity, it suffices to prove that $\Delta'(\mathbf{d})-\Delta(\mathbf{d})>0.$
  We shall prove the following three inequalities for \(\mathbf{d}\triangleleft \mathbf{d}''\) and \(\mathbf{d}'\triangleleft \mathbf{d}''\) in \S\ref{sec:Bounds-Delta}--\S\ref{sec:comparison-bounds-strict}:
  \begin{align}
    \Delta'(\mathbf{d}) & \geq \frac{1}{16}(\Delta')^{[2],\star}(\mathbf{d}), \label{eq:DeltaPrime-lower}\\ 
    \Delta(\mathbf{d}) & < \frac{1}{8}\Delta ^{[2],\star}(\mathbf{d}), \label{eq:Delta-Upper}\\ 
    (\Delta')^{[2],\star}(\mathbf{d}) & \geq 2\Delta ^{[2],\star}(\mathbf{d}). \label{eq:DeltaStar-Comp}
  \end{align}
  Combining these estimates,
  we have
  \begin{equation}\label{eq:DeltaPrime>Delta}
    \Delta'(\mathbf{d})\geq\frac{1}{16}(\Delta')^{[2],\star}(\mathbf{d})\geq\frac{1}{8}\Delta^{[2],\star}(\mathbf{d})>\Delta(\mathbf{d}).
  \end{equation}
  Together with \eqref{eq:C-B-I}, this proves \(C(\mathbf{d})<C(\mathbf{d}')\) and completes the proof.
\end{proof}

\subsection{Proof of inequalities \eqref{eq:DeltaPrime-lower} and \eqref{eq:Delta-Upper}}\label{sec:Bounds-Delta}

In this subsection, we prove \eqref{eq:DeltaPrime-lower} and \eqref{eq:Delta-Upper}. We first establish uniform bounds for the normalized BGW correlators.
\begin{theorem}\label{thm:bound-C}
  For every nonempty partition \(\mathbf{d}\), we have
  \begin{equation}
    \frac{1}{4}\leq C(\mathbf{d}) < \frac{1}{\pi}.
  \end{equation}
\end{theorem}
\begin{proof}
  By the string equation \(C(0,\mathbf{d})=C(\mathbf{d})\), it suffices to consider positive partitions and the partition \((0)\). Set \(g=|\mathbf{d}|+1\). By Corollary~\ref{cor:Bound-BGW}, it remains to prove
  \[
    C(g-1)\geq\frac{1}{4},\qquad C(1^{g-1})<\frac{1}{\pi}.
  \]
  
  We first prove the lower bound. 
  The one-point BGW correlators have the explicit formula \cite{DYZ21,BR19}
  \begin{equation}
    C(g-1)= \frac{g}{4^{2g-1}}\binom{2g-1}{g}^{2}.
  \end{equation}
  Then the sequence \(\{ C(g-1) \}_{g\in \mathbb{N}_{+}}\) is strictly increasing, since
  \[
    \frac{C(g)}{C(g-1)}=\frac{(2g+1)^{2}}{4g(g+1)}>1.
  \]
  As a consequence,
  \begin{equation}\label{eq:uniform-lower-C}
    C(\mathbf{d})\geq C(|\mathbf{d}|)\geq C(0)=\frac{1}{4}.
  \end{equation}
  We now turn to the upper bound. From \cite[(12),(15)]{GNYZ25}, we have
  \[
    C(1^{g-1})= \frac{3^{g-1}(g-1)!}{(3g-2)!}y_{g},\qquad g\geq2,
  \]
  where the sequence \(\{ y_{g} \}_{g\geq1}\) is determined by \(y_{1}=1 /4\) and the recursion
  \[
    y_{g}=(3g-2)(3g-4)y_{g-1}+\frac{2}{g-1}\sum_{k=1}^{g-1}(3k-1)y_{k}y_{g-k},\qquad g\geq2.
  \]
  The recurrence implies inductively that \(y_{g}>0\) for every \(g\in \mathbb{N}_{+}\). Substituting it into the formula for \(C(1^{g-1})\), we obtain
  \[
    C(1^{g-1})-C(1^{g-2})= \frac{2\cdot3^{g-1}(g-2)!}{(3g-2)!}\sum_{k=1}^{g-1}(3k-1)y_{k}y_{g-k}>0,\qquad g\geq3.
  \]
  Thus \(C(1^{g-1})\) is strictly increasing in \(g\). Hence, by \cite[Theorem~1]{GNYZ25}, the supremum of \(C(1^{g-1})\) is 
  \[
    \lim_{ g \to +\infty } C(1^{g-1})=\frac{1}{\pi}.
  \]
\end{proof}

We next compare the quadratic and cubic coefficient sums $K_{i,j}^{[2],\star}(\mathbf{a})$ and $K_{i,j}^{[3],\star}(\mathbf{a})$ defined in equations \eqref{eq:def-K2star} and \eqref{eq:def-K3star}, respectively.

\begin{lemma}\label{lem:cubic-absorption-K}
  Let \(\mathbf{a}=(a_{1},\dots,a_{l})\) be a positive partition, and let \(i,j \in [l]\) be distinct indices with \(a_{i},a_{j}\geq 2\). 
  We have
  \begin{equation}
    -K_{i,j}^{[3],\star}(\mathbf{a})\leq -\frac{1}{6}K_{i,j}^{[2],\star}(\mathbf{a}).
  \end{equation}
\end{lemma}
\begin{proof}
  By the definition of \(K_{i,j}^{[3],\star}\) in \eqref{eq:def-K3star}, we have 
  \[
    -K_{i,j}^{[3],\star}(\mathbf{a}) 
    =\sum_{I\sqcup J=[l]\setminus \{ i,j \}} \kappa_{(0,a_{i}-1,\mathbf{a}_{I})}\kappa_{(0,a_{j}-1,\mathbf{a}_{J})}\big(H_{\mathbf{a}_{I}}(a_{i})+H_{\mathbf{a}_{J}}(a_{j})\big),
  \]
  where we use the notation
  \begin{align}
    H_{\mathbf{a}}(m):=\frac{\beta ^{\star}_{\mathbf{a}}(m)}{\kappa_{(0,m-1,\mathbf{a})}}=\sum_{I \sqcup J=[l(\mathbf{a})]}\frac{2+\chi(\mathbf{a}_{I})}{2m+2+\chi(\mathbf{a})}\cdot \frac{\kappa_{(\mathbf{a}_{I},0^{2})}\kappa_{(\mathbf{a}_{J},0,m-1)}}{\kappa_{(\mathbf{a},0,m-1)}}, \label{eq:def-h}
  \end{align}
  for any positive partition \(\mathbf{a}\) and integer \(m\geq2\). We claim that
  \begin{equation}\label{eq:estimate-h}
    H_{\mathbf{a}}(m)\leq \frac{\chi(m,\mathbf{a})-1}{24}.
  \end{equation}
  Using this claim, we have 
  \begin{align*}
    -K_{i,j}^{[3],\star}(\mathbf{a}) & \leq \sum_{I\sqcup J=[l]\setminus \{ i,j \}}\frac{\chi(a_{i},\mathbf{a}_{I})+\chi(a_{j},\mathbf{a}_{J})-2}{24}\kappa_{(0,a_{i}-1,\mathbf{a}_{I})}\kappa_{(0,a_{j}-1,\mathbf{a}_{J})} \\
    & \leq \sum_{I\sqcup J=[l]\setminus \{ i,j \}} \frac{Z(a_{i},a_{j},\mathbf{a}_{I},\mathbf{a}_{J})}{6}\kappa_{(0,a_{i}-1,\mathbf{a}_{I})}\kappa_{(0,a_{j}-1,\mathbf{a}_{J})} 
    = -\frac{1}{6}K_{i,j}^{[2],\star}(\mathbf{a}),
  \end{align*}
which completes the proof of this Lemma.

  To prove the estimate \eqref{eq:estimate-h}, we first derive an integral representation for \(H_{\mathbf{a}}(m)\) by using the beta integral.
  By the definition of $H_{\mathbf{a}}(m)$, we have
  \begin{align}\label{eqn:H as int}
  	\begin{split}
    H_{\mathbf{a}}(m) & =\frac{1}{2}\sum_{I \sqcup J=[l(\mathbf{a})]}\frac{2+\chi(\mathbf{a}_{I})}{2m+2+\chi(\mathbf{a})}\cdot \frac{(\chi(\mathbf{a}_{I})+1)!(\chi(\mathbf{a}_{J})+2m-1)!}{(\chi(\mathbf{a})+2m-1)!} \\
    & = \frac{(2m+1+\chi(\mathbf{a}))(2m+\chi(\mathbf{a}))}{2}\sum_{I\sqcup J=[l(\mathbf{a})]}\int_{0}^{1}t^{2+\chi(\mathbf{a}_{I})}(1-t)^{2m-1+\chi(\mathbf{a}_{J})}\,\dd t \\
    & =\frac{\chi(m,\mathbf{a})(\chi(m,\mathbf{a})-1)}{2}\int_{0}^{1}t^{2}(1-t)^{2m-1}\mathcal{T}_{\mathbf{a}}(t)\,\dd t,
    \end{split}
  \end{align}
  where
  \begin{equation}\label{eqb:def T}
    T_{j}(t)=t^{j}+(1-t)^{j},\quad\mathcal{T}_{\xi}(t)=\prod_{i=1}^{l(\xi)}T_{2\xi_{i}+1}(t),
  \end{equation}
  for \(j\in \mathbb{N}\) and any partition \(\xi=(\xi_1,\dots,\xi_{l(\xi)})\). Since \(\mathcal{T}_{\mathbf{a}}(1-t)=\mathcal{T}_{\mathbf{a}}(t)\), the change of variables \(t\mapsto1-t\) gives a second representation of $H_{\mathbf{a}}(m)$ as
  \begin{align*}
    H_{\mathbf{a}}(m)=\frac{\chi(m,\mathbf{a})(\chi(m,\mathbf{a})-1)}{2}\int_{0}^{1}t^{2m-1}(1-t)^{2}\mathcal{T}_{\mathbf{a}}(t)\,\dd t.
  \end{align*}
  Averaging the above two representations yields
  \begin{equation}\label{eq:h-beta-integral}
    H_{\mathbf{a}}(m)=\frac{\chi(m,\mathbf{a})(\chi(m,\mathbf{a})-1)}{4}\int_{0}^{1}t^{2}(1-t)^{2}\mathcal{T}_{(m-2,\mathbf{a})}(t)\,\dd t.
  \end{equation}

  Based on the above formula \eqref{eq:h-beta-integral},
  to estimate $H_{\mathbf{a}}(m)$, 
  we need to study the function \(\mathcal{T}_{\xi}(t)\).
  Note that for any positive partition \(\xi\) and \(x \in \xi\), we have
  \begin{equation*}
    t^{2x+1}+(1-t)^{2x+1} \leq (t^{3}+(1-t)^{3})^{\frac{2x+1}{3}}\leq (e^{-3t(1-t)})^{\frac{2x+1}{3}}=e^{-(2x+1)t(1-t)}.
  \end{equation*}
  Then from the definition \eqref{eqb:def T} of $\mathcal{T}_{\xi}(t)$, we have
  \begin{equation}\label{eq:estimate-T-exp}
    \mathcal{T}_{\xi}(t)\leq \exp\big(-\chi(\xi)t(1-t)\big)
    \leq \frac{1}{t(1-t)} \frac{1}{e \cdot \chi(\xi)}.
  \end{equation}
  
  Now,
  we prove the estimate \eqref{eq:estimate-h}.
  For the case of \(m=2\) and \(\mathbf{a}=\emptyset\), we have 
  \begin{align}\label{eqn:m=2a=emp}
  	H_{\emptyset}(2)= \frac{5\cdot 4}{4}\int_{0}^{1}t^{2}(1-t)^{2}\,\dd t = \frac{1}{6}=\frac{\chi(2)-1}{24}.
  \end{align}
  For the case of \(m=2\) and \(\mathbf{a}\neq \emptyset\), since \(\chi(\mathbf{a})\geq3\), applying \eqref{eq:estimate-T-exp} to \eqref{eq:h-beta-integral} gives
  \begin{align}\label{eqn:m=2,a}
  	\begin{split}
    H_{\mathbf{a}}(2)
    & \leq \frac{(5+\chi(\mathbf{a}))(4+\chi(\mathbf{a}))}{4}\int_{0}^{1}\frac{t(1-t)}{\chi(\mathbf{a})\cdot e}\,\dd t  \\
    & =\frac{5+\chi(\mathbf{a})}{e\cdot \chi(\mathbf{a})}\cdot \frac{\chi(2,\mathbf{a})-1}{24} 
     \leq \frac{8}{3e}\cdot\frac{\chi(2,\mathbf{a})-1}{24} 
    \leq\frac{\chi(2,\mathbf{a})-1}{24} .
    \end{split}
  \end{align}
   The proof of the case of \(m\geq3\) is similar to the previous case.
   Since \(\chi(m-2,\mathbf{a})\geq3\), we have
  \begin{align}\label{eqn:m=3 case}
  	\begin{split}
    H_{\mathbf{a}}(m)
    & \leq \frac{\chi(m,\mathbf{a})(\chi(m,\mathbf{a})-1)}{4}\int_{0}^{1} \frac{t(1-t)}{e\cdot\chi(m-2,\mathbf{a})}\,\dd t \\
    & = \frac{4+\chi(m-2,\mathbf{a})}{e\cdot \chi(m-2,\mathbf{a})}\cdot\frac{\chi(m,\mathbf{a})-1}{24} 
     \leq \frac{7}{3e}\cdot\frac{\chi(m,\mathbf{a})-1}{24} 
     \leq \frac{\chi(m,\mathbf{a})-1}{24}.
     \end{split}
  \end{align}
  The above three formulas \eqref{eqn:m=2a=emp}-\eqref{eqn:m=3 case} prove the estimate \eqref{eq:estimate-h} for $H_{\mathbf{a}}(m)$ in all cases and complete the proof of the lemma.
\end{proof}

In the following,
we prove inequalities \eqref{eq:DeltaPrime-lower} and \eqref{eq:Delta-Upper}.
For convenience,
we restate them as the following proposition.
\begin{proposition}\label{prp:Delta-bound}
  Let \(\mathbf{d},\mathbf{d}'\) be partitions satisfying the assumptions of Theorem~\ref{thm:equal-length}, and let the auxiliary partition \(\mathbf{d}''\) satisfy \(\mathbf{d}''\triangleright \mathbf{d}\). 
  Using the notations of \eqref{eq:alpha-j}, \eqref{eq:alpha-j-star} and \eqref{eq:def-Delta-Delta'}, we have
  \begin{equation}\label{eq:Delta-Bound}
    \Delta'(\mathbf{d})\geq\frac{1}{16}(\Delta')^{[2],\star}(\mathbf{d}),\quad\Delta(\mathbf{d})< \frac{1}{8}\Delta ^{[2],\star}(\mathbf{d}).
  \end{equation}
\end{proposition}
\begin{proof}
  Recall that \((\Delta')^{[2]}(\mathbf{d})\) is a sum of products of two normalized BGW correlators with nonnegative coefficients. The sum of these coefficients is \((\Delta')^{[2],\star}(\mathbf{d})\). 
  By Theorem~\ref{thm:bound-C}, each normalized BGW correlator is at least \(1/4\). Since \((\Delta')^{[3]}(\mathbf{d})\) is nonnegative, we obtain 
  \[
    \Delta'(\mathbf{d})=(\Delta')^{[2]}(\mathbf{d})+(\Delta')^{[3]}(\mathbf{d})\geq(\Delta')^{[2]}(\mathbf{d})\geq \frac{1}{16}(\Delta')^{[2],\star}(\mathbf{d}).
  \]
  This proves the first inequality of \eqref{eq:Delta-Bound}.

  Along the transfer chain \eqref{eq:d-transfer-chain} from \(\mathbf{d}''\) to \(\mathbf{d}\), 
  note that \(d_{k_{j}}^{(j)},d_{m_{j}}^{(j)}\geq2\) for all \(j=1,\dots,q\),
  then by using the notation in \eqref{eq:alpha-j-star} and Lemma \ref{lem:cubic-absorption-K},
  we have
  \begin{align}
    \Delta ^{[2],\star}(\mathbf{d}) & = -\frac{1}{2}\sum_{j=1}^{q}\sum_{k=0}^{d_{m_{j}}^{(j)}-d_{k_{j}}^{(j)}}K^{[2],\star}_{k_{j},m_{j}}(\mathbf{d}^{(j)}+k(e_{k_{j}}-e_{m_{j}})) \label{eq:Delta2star-ineq} \\
    & \geq -3\sum_{j=1}^{q}\sum_{k=0}^{d_{m_{j}}^{(j)}-d_{k_{j}}^{(j)}}K^{[3],\star}_{k_{j},m_{j}}(\mathbf{d}^{(j)}+k(e_{k_{j}}-e_{m_{j}})) 
     =6\Delta ^{[3],\star}(\mathbf{d}). \label{eq:Delta3star}
  \end{align}
  By Theorem~\ref{thm:bound-C}, each normalized BGW correlator is at most \(1/\pi\),
  so we have
  \begin{equation*}
    \Delta(\mathbf{d})=\Delta ^{[2]}(\mathbf{d})+\Delta ^{[3]}(\mathbf{d})
    \leq\frac{1}{\pi^2}\Delta ^{[2],\star}(\mathbf{d})+\frac{1}{\pi^3}\Delta ^{[3],\star}(\mathbf{d})
    \leq \left(\frac{1}{\pi^2} +\frac{1}{6\pi^3}\right)\Delta ^{[2],\star}(\mathbf{d})<\frac{1}{8}\Delta ^{[2],\star}(\mathbf{d}).
  \end{equation*}
      Here the last inequality is strict, since \(\Delta^{[2],\star}(\mathbf{d})>0\) by \(\mathbf{d}''\triangleright\mathbf{d}\).
 This proves the second inequality of \eqref{eq:Delta-Bound}.

\end{proof}

\subsection{Proof of the inequality \eqref{eq:DeltaStar-Comp}}\label{sec:comparison-bounds-strict}  
In this subsection, we derive integral formulas for \(\Delta ^{[2],\star}(\mathbf{d})\) and \((\Delta') ^{[2],\star}(\mathbf{d})\), and then use them to prove the inequality \eqref{eq:DeltaStar-Comp}.
\begin{lemma}\label{lem:Delta2Star-Computation}
  Let \(\mathbf{a},\mathbf{b}\in \mathcal{P}_{s,l}\) be nonempty partitions satisfying \(\mathbf{a}\trianglerighteq \mathbf{b}\). Then 
  \begin{equation}
    (B(\mathbf{b})-B(\mathbf{a}))^{[2],\star}=\kappa_{\mathbf{a}}\int_{0}^{1}(\mathcal{T}_{\mathbf{a}}(t)-\mathcal{T}_{\mathbf{b}}(t))\dd t.
  \end{equation}
\end{lemma}

\begin{proof}
	By Lemma~\ref{lem:muirhead},
	there exists a transfer chain from $\mathbf{a}$ to $\mathbf{b}$.
	Then from the linearity and the fact that transfers preserve the normalization factor \(\kappa_{\mathbf{d}}\),  we only need to deal with the case of a single transfer,
	i.e. \(\mathbf{b}=R_{i,j}(\mathbf{a})\) for one transfer \(R_{i,j}\). By \eqref{eq:alpha-j-star}, in this case, we have
  \begin{equation}\label{eq:B2-K2}
    (B(\mathbf{b})-B(\mathbf{a}))^{[2],\star}  =-\frac{1}{2}\sum_{k=0}^{a_{j}-a_{i}-2}K_{i,j}^{[2],\star}\big(\mathbf{a}+(k+1)(e_{i}-e_{j})\big).
  \end{equation}
  To compute the right hand side of the above equation, we first derive an integral representation for \(K_{i,j}^{[2],\star}(\mathbf{c})\) for arbitrary \(\mathbf{c}\) with \(c_{i},c_{j}>0\). 
  Recall that the \(K_{i,j}^{[2],\star}(\mathbf{c})\) is given by \eqref{eq:def-K2star},
  then using the similar method in deriving the integral expression \eqref{eqn:H as int}
   of $H_{\mathbf{a}}(m)$, we have
  \begin{equation}\label{eq:K2star-intT}
    \begin{aligned}
      -K_{i,j}^{[2],\star}(\mathbf{c}) & =\sum_{I\sqcup J=[l(\mathbf{c})]\setminus \{ i,j \}}
      \frac{\big(\chi(c_{i},\mathbf{c}_{I})+\chi(c_{j},\mathbf{c}_{J})-1\big)!}{2^{2|\mathbf{c}|}}\int_{0}^{1}t^{\chi(c_{i},\mathbf{c}_{I})-2}(1-t)^{\chi(c_{j},\mathbf{c}_{J})-2}(1-2t)^{2}\,\dd t \\
      & =2\kappa_{\mathbf{c}}\int_{0}^{1}t^{2c_{i}-1}(1-t)^{2c_{j}-1}(1-2t)^{2}\mathcal{T}_{\mathbf{c}_{[l(\mathbf{c})]\setminus \{ i,j \}}}(t)\,\dd t,
    \end{aligned}
  \end{equation}
  where we recall that $\mathcal{T}_{\xi}(t)=\prod_{i=1}^{l(\xi)}\big(t^{2\xi_{i}+1}+(1-t)^{2\xi_{i}+1}\big)$.
  We apply the above representation of $K_{i,j}^{[2],\star}(\mathbf{c})$ at \(\mathbf{c}=\mathbf{a}+(k+1)(e_{i}-e_{j})\), which satisfies that the $i$-th and \(j\)-th parts are positive for \(0\leq k\leq a_{j}-a_{i}-2\), even when \(a_{i}=0\). 
  To sum the terms in the right hand side of equation \eqref{eq:B2-K2}, observe that
  \begin{equation}\label{eq:sum-TT}
    (1-2t)^{2}\sum_{k=0}^{a_{j}-a_{i}-2}t^{2a_{i}+2k+1}(1-t)^{2a_{j}-2k-3}= T_{2a_{i}+1}(t)T_{2a_{j}+1}(t)-T_{2a_{i}+3}(t)T_{2a_{j}-1}(t).
  \end{equation}
  Then by \eqref{eq:B2-K2}–\eqref{eq:sum-TT}, we have 
  \begin{align*}
    (B(\mathbf{b})-B(\mathbf{a}))^{[2],\star}
    = \kappa_{\mathbf{a}}\int_{0}^{1}(\mathcal{T}_{\mathbf{a}}(t)-\mathcal{T}_{\mathbf{b}}(t))\,\dd t.
  \end{align*}

\end{proof}

In the following,
we prove the inequality \eqref{eq:DeltaStar-Comp}.
We restate it as the following proposition.
\begin{proposition}\label{prp:quadratic-Delta-double}
  Let \(\mathbf{d},\mathbf{d}'\) be partitions satisfying the assumptions of Theorem~\ref{thm:equal-length}. With the notation in \eqref{eq:alpha-j-star} and \eqref{eq:def-Delta-Delta'}, we have
  \begin{equation}
    (\Delta')^{[2],\star}(\mathbf{d})\geq 2\Delta ^{[2],\star}(\mathbf{d}).
  \end{equation}
\end{proposition}
\begin{proof}
  We retain the notation \(l=l(\mathbf{d}),m=m(\mathbf{d}),n=l-m\) and \(\pi=(d_{1},\dots,d_{m-1})\) for the common prefix of \(\mathbf{d},\mathbf{d}',\mathbf{d}''\) as introduced in the proof of Theorem~\ref{thm:equal-length}. Denote the corresponding tails of \(\mathbf{d},\mathbf{d}',\mathbf{d}''\) by \(\rho,\rho',\rho''\), respectively. By Lemma~\ref{lem:Delta2Star-Computation} and the fact that \(T_{m}(t)\) is symmetric with respect to \(t=\frac{1}{2}\) for all \(m \in \mathbb{N}\), we have
  \begin{align}
    (\Delta')^{[2],\star}(\mathbf{d}) & =2\kappa_{\mathbf{d}} \int_{0}^{\frac{1}{2}}\mathcal{T}_{\pi}(t)(\mathcal{T}_{\rho''}(t)-\mathcal{T}_{\rho'}(t))\,\dd t ,\label{eq:Delta2Prime-Star} \\
    \Delta ^{[2],\star}(\mathbf{d}) & =2\kappa_{\mathbf{d}} \int_{0}^{\frac{1}{2}}\mathcal{T}_{\pi}(t)(\mathcal{T}_{\rho''}(t)-\mathcal{T}_{\rho}(t))\,\dd t,\label{eq:Delta2-Star}
  \end{align}
  where we have used the property $\mathcal{T}_{(\pi,\rho)}(t)=\mathcal{T}_{\pi}(t)\cdot \mathcal{T}_{\rho}(t)$.
  Then throughout the remainder of the proof, we can assume that \(t \in\left[0,\frac{1}{2}\right]\). 

  For convenience, denote
  \[
    a:=d_{m},\qquad c:=n(d_{m+1}-d_{m}-1)+d_{m}+r.
  \]
  Then we have
  \[
    \rho'=((a+1)^{n},c),\qquad \rho''=(a,(a+1)^{n-1},c+1).
  \]
  
  To compare the two integral representations in \eqref{eq:Delta2Prime-Star} and \eqref{eq:Delta2-Star},
  we derive a formula for \(\mathcal{T}_{\rho''}(t)-\mathcal{T}_{\rho'}(t)\) and an estimate for \(\mathcal{T}_{\rho''}(t)-\mathcal{T}_{\rho}(t)\) in Lemma \ref{eqn:Trho1} and Lemma \ref{lem:T estimate}, respectively,
  \begin{align}
  	\mathcal{T}_{\rho''}(t)-\mathcal{T}_{\rho'}(t)  =&T_{2a+3}(t)^{n-1} \cdot S(t)^{2a+1}R_{2}(t)R_{2(c-a)}(t),\label{eq:difference-rho2-rho1}\\
  	\mathcal{T}_{\rho''}(t) -\mathcal{T}_{\rho}(t)
  	\leq& (n-1)T_{2a+3}(t)^{n-2}\cdot T_{2a+1}(t)S(t)^{2a+3}R_{2(c-a)}(t), \label{eq:uppper-difference-T-eta2p-eta}
  \end{align}
  where we have used the notation $R_{j}(t)=(1-t)^{j}-t^{j}$ and $S(t)=t(1-t)$.
  Then we have
  \begin{align}\label{eqn:TTTT}
  	\begin{split}
  	&\big(\mathcal{T}_{\rho''}(t)-\mathcal{T}_{\rho'}(t) \big)
  	-2\big(\mathcal{T}_{\rho''}(t) -\mathcal{T}_{\rho}(t)\big)\\
  	&\ \ \ \ \geq
  	T_{2a+3}(t)^{n-1} S(t)^{2a+1}R_{2}(t)R_{2(c-a)}(t)
  	- 2(n-1)T_{2a+3}(t)^{n-2} T_{2a+1}(t)S(t)^{2a+3}R_{2(c-a)}(t),
  	\end{split}
  \end{align}

  We introduce an auxiliary function $U(t)= V(t)W(t)$,
  where
  \[
  V(t)=\frac{S(t)^{2a+2}}{2a+3}\left(R_{2}(t)+\frac{1}{2a+2}\right),\quad W(t)=\frac{R_{2(c-a)}(t)}{R_{2}(t)}.
  \]
  Next,
  we prove three properties \eqref{eq:U-cdt1}, \eqref{eq:U-cdt3} and \eqref{eq:U-cdt2} of the function $U(t)$. 
  Note that
  \begin{align}\label{eq:W-expansion}
  	\begin{split}
  	W(t) & = \frac{1}{R_{2}(t)}\bigg(\Big( \frac{1+R_{2}(t)}{2} \Big) ^{2(c-a)}-\Big( \frac{1-R_{2}(t)}{2} \Big) ^{2(c-a)}\bigg) \\
  	& =2^{1-2(c-a)}\sum_{k=0}^{c-a-1}\binom{2(c-a)}{2k+1}R_{2}(t)^{2k}>0, 
  	\end{split}
  \end{align}
  so, together with $V(t)\geq0$,
  the auxiliary function \(U(t)\) satisfies
  \begin{equation}\label{eq:U-cdt1}
  	U(0)=0,\qquad U(t)\geq0,\quad t\in\left[0,\frac{1}{2}\right],
  \end{equation}
  Moreover,
  by taking the derivative of \eqref{eq:W-expansion}, we obtain
  \[
  W'(t)=-2^{3-2(c-a)}\sum_{k=1}^{c-a-1}k\binom{2(c-a)}{2k+1}R_{2}(t)^{2k-1}\leq0.
  \]
  Similarly,
  the derivative of $V(t)$ is given by
  \[
  V'(t)=S(t)^{2a+1}R_{2}(t)^{2}- \frac{2}{2a+3}S(t)^{2a+1}t^{2}.
  \]
  Since \(V(t)\geq0\) and \(W'(t)\leq0\), we have
  \begin{align}\label{eq:U-cdt3}
  	\begin{split}
  	U'(t) & =V'(t)W(t)+V(t)W'(t) 
  	 \leq V'(t)W(t) \\
  	& =S(t)^{2a+1}R_{2}(t)R_{2(c-a)}(t)- \frac{2}{2a+3}S(t)^{2a+1}t^{2}W(t) 
  	 \leq S(t)^{2a+1}R_{2}(t)R_{2(c-a)}(t).
  	 \end{split}
  \end{align}
  By $(T_{2a+3}(t)^{n-1})'=-(n-1)(2a+3)T_{2a+3}(t)^{n-2}R_{2a+2}(t)$ and the definition of $U(t)$,
  we have
  \begin{align}\label{eq:U-cdt2}
  	\begin{split}
  	-(T_{2a+3}(t)^{n-1})'U(t) & =(n-1)T_{2a+3}(t)^{n-2}
  	R_{2(c-a)}(t)S(t)^{2a+2}
  	\Big(R_{2a+2}(t)+ \frac{R_{2a+2}(t)}{(2a+2)R_{2}(t)}\Big) \\
  	& \geq(n-1)T_{2a+3}(t)^{n-2}R_{2(c-a)}(t)S(t)^{2a+2}\left(R_{2a+2}(t)+t^{2a+1}\right) \\
  	& =(n-1)T_{2a+3}(t)^{n-2}R_{2(c-a)}(t)S(t)^{2a+2}(1-t)T_{2a+1}(t) \\
  	& \geq2(n-1)T_{2a+3}(t)^{n-2}R_{2(c-a)}(t)S(t)^{2a+3}T_{2a+1}(t),
  	\end{split}
  \end{align}
  where in the last line we have used $1-t \geq(1-t)\cdot 2t=2S(t)$,
  and in the second line we have used
  \begin{align*}
  	\frac{R_{2a+2}(t)}{R_{2}(t)} =\frac{R_{2a+2}(t)}{R_{1}(t)}=\sum_{k=0}^{2a+1}(1-t)^{k}t^{2a+1-k}\geq2(a+1)t^{2a+1}.
  \end{align*}
  
  Finally, by using the properties \eqref{eq:U-cdt3}, \eqref{eq:U-cdt2} of $U(t)$,
  the formula \eqref{eqn:TTTT} gives
  \begin{align*}
   (\mathcal{T}_{\rho''}(t)-\mathcal{T}_{\rho'}(t))-2(\mathcal{T}_{\rho''}(t)-\mathcal{T}_{\rho}(t)) 
    &\geq U'(t)T_{2a+3}(t)^{n-1}+U(t)\big(T_{2a+3}(t)^{n-1}\big)' \\
    &= \big(U(t)T_{2a+3}(t)^{n-1}\big)'.
  \end{align*}
  Then by the integral representations \eqref{eq:Delta2Prime-Star} and \eqref{eq:Delta2-Star} for 
  $(\Delta')^{[2],\star}(\mathbf{d})$ and 
  $\Delta ^{[2],\star}(\mathbf{d})$,
  we obtain
  \begin{align*}
    (\Delta')^{[2],\star}(\mathbf{d})-&2\Delta ^{[2],\star}(\mathbf{d}) \geq2\kappa_{\mathbf{d}} \int_{0}^{\frac{1}{2}}\mathcal{T}_{\pi}(t)\big(U(t)T_{2a+3}(t)^{n-1}\big)'\,\dd t \\
    & =2\kappa_{\mathbf{d}}
    \bigg(\mathcal{T}_{\pi}\Big(\frac{1}{2}\Big)
    U\Big(\frac{1}{2}\Big)
    T_{2a+3}\Big(\frac{1}{2}\Big)^{n-1}
    -\int_{0}^{\frac{1}{2}}\mathcal{T}_{\pi}'(t)U(t)T_{2a+3}(t)^{n-1}\,\dd t\bigg)
    \geq 0.
  \end{align*}
  where we have used $U(0)=0, U(t)\geq0$, and \(\mathcal{T}_{\pi}'(t)\leq0\) for \(t \in[0,1 /2 ]\).
  This completes the proof.
\end{proof}

In the following lemmas,
we study $\mathcal{T}_{\rho''}(t)-\mathcal{T}_{\rho'}(t)$ and $\mathcal{T}_{\rho''}(t)-\mathcal{T}_{\rho}(t)$,
which have been used in the proof of Proposition \ref{prp:quadratic-Delta-double}.
\begin{lemma}\label{eqn:Trho1}
	We have
	\begin{align}
		\mathcal{T}_{\rho''}(t)-\mathcal{T}_{\rho'}(t)  =T_{2a+3}(t)^{n-1} \cdot S(t)^{2a+1}R_{2}(t)R_{2(c-a)}(t).
		\label{eq:difference-rho2-rho1 lemma}
	\end{align}
\end{lemma}
\begin{proof}
	 By the notation $R_{j}(t)=(1-t)^{j}-t^{j}, S(t)=t(1-t)$,
	the function $T_j(t)$ satisfies the following  identity
	\begin{equation}\label{eq:TT-SRR}
		T_{p}(t)T_{p+q+r}(t)=T_{p+q}(t)T_{p+r}(t)+S(t)^{p}R_{q}(t)R_{r}(t)
	\end{equation}
	for all \(p,q,r\in \mathbb{N}\).
	Then by taking \(p=2a+1,q=2,r=2(c-a)\) in \eqref{eq:TT-SRR},
	we obtain this lemma.
\end{proof}

\begin{lemma}\label{eqn:Trho2}
	We have
	\begin{equation}\label{eq:T-etaprime2-eta lemma}
		\mathcal{T}_{\rho''}(t) -\mathcal{T}_{\rho}(t)=T_{2a+1}(t)S(t)^{2a+3}\sum_{j=2}^{n}\bigg(T_{2a+3}(t)^{j-2}R_{2E_{j-1}}(t)R_{2\varepsilon_{j}}(t)\prod_{k=j+1}^{n}T_{2a+3+2\varepsilon_{k}}(t)\bigg).
	\end{equation}
\end{lemma}
\begin{proof}
	Set \(\varepsilon=(\varepsilon_{1},\dots,\varepsilon_{n})\) with \(\varepsilon_{i}=d_{m+i}-(a+1)\). Let \(E_{j}=\sum_{i=1}^{j}\varepsilon_{i}\). Then \( c-a=|\varepsilon|\). 
	Applying the formula \eqref{eq:TT-SRR} successively to \(\mathcal{T}_{\rho''}(t)\) yields
	\begin{align*}
		\mathcal{T}_{\rho''}(t)
		=& T_{2a+1}(t)T_{2a+3}(t)^{n-1}T_{2c+3}(t)  \\
		=& T_{2a+1}(t)T_{2a+3}(t)^{n-2}\big(T_{2a+3}(t)T_{2a+3+2|\varepsilon|}(t)\big) \\
		=& T_{2a+1}(t)\Big( T_{2a+3}(t)^{n-2}T_{2a+3+2E_{n-1}}(t)T_{2a+3+2\varepsilon_{n}}(t) \\
		& \qquad\qquad +T_{2a+3}(t)^{n-2}S(t)^{2a+3}R_{2E_{n-1}}(t)R_{2\varepsilon_{n}}(t) \Big)  \\
		=& T_{2a+1}(t)\Big(T_{2a+3}(t)^{n-3}T_{2a+3+2E_{n-2}}(t)T_{2a+3+2\varepsilon_{n-1}}(t)T_{2a+3+2\varepsilon_{n}}(t) \\
		& \qquad\qquad +S(t)^{2a+3}T_{2a+3}(t)^{n-3}R_{2E_{n-2}}(t)R_{2\varepsilon_{n-1}}(t)T_{2a+3+2\varepsilon_{n}}(t) \\
		& \qquad\qquad +S(t)^{2a+3}T_{2a+3}(t)^{n-2}R_{2E_{n-1}}(t)R_{2\varepsilon_{n}}(t)\Big) \\
		=& \cdots \\
		=& T_{2a+1}(t)\Bigg(\prod_{i=1}^{n}T_{2a+3+2\varepsilon_{i}}(t) \\
		& \qquad\qquad +S(t)^{2a+3}\sum_{j=2}^{n}T_{2a+3}(t)^{j-2}R_{2E_{j-1}}(t)R_{2\varepsilon_{j}}(t)\prod_{k=j+1}^{n}T_{2a+3+2\varepsilon_{k}}(t)\Bigg),
	\end{align*}
	which yields \eqref{eq:T-etaprime2-eta lemma} since $\mathcal{T}_{\rho}(t)
	=T_{2a+1}(t)\prod_{i=1}^{n}T_{2d_{m+i}+1}(t)=T_{2a+1}(t)
	\prod_{i=1}^{n}T_{2a+3+2\varepsilon_{i}}(t)$.
\end{proof}

\begin{lemma}\label{lem:T estimate}
	For \(t \in[0,1/ 2]\),
	we have the following estimate
	\begin{equation}\label{eq:uppper-difference-T-eta2p-eta lemma}
		\mathcal{T}_{\rho''}(t) -\mathcal{T}_{\rho}(t)\leq(n-1)T_{2a+3}(t)^{n-2}\cdot T_{2a+1}(t)S(t)^{2a+3}R_{2(c-a)}(t).
	\end{equation}
\end{lemma}
\begin{proof}
	We have
	\begin{align*}
		T_{p+q}(t) & =t^{p+q}+(1-t)^{p+q}\leq t^{p}(1-t)^{q}+(1-t)^{p+q}=(1-t)^{q}T_{p}(t), \\
		R_{p}(t)R_{q}(t) & =((1-t)^{p+q}-t^{p+q})-(t^{p}(1-t)^{q}+t^{q}(1-t)^{p}-2t^{p+q})\leq R_{p+q}(t), \\
		(1-t)^{p}R_{q}(t) & =(1-t)^{p+q}-(1-t)^{p}t^{q}\leq (1-t)^{p+q}-t^{p+q}=R_{p+q}(t).
	\end{align*}
	Applying these three inequalities to the summand in \eqref{eq:T-etaprime2-eta lemma}, one obtains
	\begin{align*}
	 T_{2a+3}(t)^{j-2}&R_{2E_{j-1}}(t)R_{2\varepsilon_{j}}(t)\prod_{k=j+1}^{n}T_{2a+3+2\varepsilon_{k}}(t) \\
		\leq & T_{2a+3}(t)^{j-2}\cdot R_{2E_{j}}(t)\cdot(1-t)^{2(|\varepsilon|-E_{j})}T_{2a+3}(t)^{n-j} 
		\leq  T_{2a+3}(t)^{n-2}R_{2|\varepsilon|}(t),
	\end{align*}
	which, together with Lemma \ref{eqn:Trho2}, directly proves \eqref{eq:uppper-difference-T-eta2p-eta lemma}.
\end{proof}

\subsection{Strict monotonicity in the different-length case}\label{sec:case-different-lengths}
In this subsection, we complete the proof of Theorem~\ref{thm:monotonicity-BGW} by treating the case of adjacent partitions of different lengths.
\begin{theorem}\label{thm:different-length}
  Let \(g\geq2\), and let \(\mathbf{d}\prec \mathbf{d}'\) be adjacent positive partitions of \(g-1\) with different lengths. 
  We have \(C(\mathbf{d})<C(\mathbf{d}')\).
\end{theorem}
\begin{proof}
   Set \(l=l(\mathbf{d})\) and write \(g-1=dl+t\) with \(0\leq t< l\). By Proposition~\ref{prop:adjacent-partitions}, the partitions \(\mathbf{d}\) and \(\mathbf{d}'\) have the following forms:
  \[
    \mathbf{d}=(d^{l-t},(d+1)^{t}),\quad \mathbf{d}'=(1^{l},g-1-l).
  \]
  Define the auxiliary partition in this case by
  \begin{equation}
    \mathbf{d}''=(1^{l-1},g-l).
  \end{equation}
  To use the method in Theorem \ref{thm:equal-length},
  we need partitions with equal length.
  So, we set
  \begin{equation}
    \rho=(0,\mathbf{d}),\quad \rho'=\mathbf{d}',\quad \rho''=(0,\mathbf{d}'').
  \end{equation}
  As a consequence,
  $\rho'$ can be obtained from $\rho''$ by a single transfer,
  i.e., $\rho'=R_{1,l+1}(\rho'')$.
  Moreover, since \(\mathbf{d}\trianglelefteq\mathbf{d}''\), we have \(\rho\trianglelefteq \rho''\) by Remark~\ref{rem:order-iso}, and hence by Lemma~\ref{lem:muirhead},
  there exists a transfer chain from $\rho''$ to $\rho$.
  Define
  \begin{equation}
    \Delta(\rho)=B(\rho)-B(\rho''),\quad \Delta'(\rho)=B(\rho')-B(\rho'').
  \end{equation}
  The string equation gives
  \begin{equation}\label{eq:C-B-II}
    C(\mathbf{d}')-C(\mathbf{d})=\frac{B(\mathbf{d}')-B(0,\mathbf{d})}{\kappa_{(0,\mathbf{d})}}=\frac{\Delta'(\rho)-\Delta(\rho)}{\kappa_{(0,\mathbf{d})}}.
  \end{equation}
  It therefore suffices to prove that
  \[
    \Delta'(\rho)-\Delta(\rho)>0.
  \]
  If \(\rho'\) can be obtained from \(\rho\) by a single transfer, 
  the desired inequality $C(\rho)<C(\rho')$ follows from the Schur-concavity in Theorem~\ref{thm:monotone-pseudo-adjacent}.
   Then we may assume that no such transfer exists. In particular, \(\rho''\neq \rho\) and \(\Delta(\rho)>0\).

  Similar to the proof of Theorem \ref{thm:equal-length},
  we need to establish three inequalities like \eqref{eq:DeltaPrime-lower}-\eqref{eq:DeltaStar-Comp}.
  The proofs of the lower bound in Proposition~\ref{prp:Delta-bound} and the comparison in Proposition~\ref{prp:quadratic-Delta-double} remain valid.
  They therefore give
  \begin{align}
    \Delta'(\rho) & \geq \frac{1}{16}(\Delta')^{[2],\star}(\rho), \label{eqn:1/16}\\
    (\Delta')^{[2],\star}(\rho) & \geq2\Delta ^{[2],\star}(\rho), \label{eqn:2}
  \end{align}
  which correspond to \eqref{eq:DeltaPrime-lower} and \eqref{eq:DeltaStar-Comp}, respectively.
  It remains to establish
  \begin{equation}\label{eq:Delta2-comp}
    \Delta(\rho)< \frac{1}{8}\Delta ^{[2],\star}(\rho),
  \end{equation}
  which corresponds to \eqref{eq:Delta-Upper}.
  To prove it,
  recall that
  \[
    \Delta(\mathbf{d})=B(\mathbf{d})-B(\mathbf{d}'').
  \]
  By Lemma~\ref{lem:Delta2Star-Computation} and the string equation, we have
  \begin{align}\label{eqn:rho-d}
  	\begin{split}
    \Delta ^{[2],\star}(\rho) & =\kappa_{(0,\mathbf{d})}\int_{0}^{1}
    \big(\mathcal{T}_{(0,\mathbf{d}'')}(t)-\mathcal{T}_{(0,\mathbf{d})}(t)\big)\,\dd t \\
    & =\chi(\mathbf{d})\kappa_{\mathbf{d}}\int_{0}^{1}(\mathcal{T}_{\mathbf{d}''}(t)-\mathcal{T}_{\mathbf{d}}(t))\,\dd t
    =\chi(\mathbf{d})\Delta ^{[2],\star}(\mathbf{d}).
    \end{split}
  \end{align}
  Since \(\mathbf{d}''\triangleright \mathbf{d}\) and both partitions are positive, fix a transfer chain from \(\mathbf{d}''\) to \(\mathbf{d}\) through positive partitions. Using this chain to define the components of \(\Delta(\mathbf{d})\), the upper-bound argument in the proof of Proposition~\ref{prp:Delta-bound} gives
  \begin{align}\label{eqn:tri<tri-8}
  	 \Delta(\mathbf{d})<\frac{1}{8}\Delta ^{[2],\star}(\mathbf{d}).
  \end{align}
  Using  the equality \eqref{eqn:rho-d} and the inequality \eqref{eqn:tri<tri-8}, we therefore obtain
  \[
    \Delta(\rho)=\chi(\mathbf{d})\Delta(\mathbf{d})< \frac{1}{8}\chi(\mathbf{d})\Delta ^{[2],\star}(\mathbf{d})=\frac{1}{8}\Delta ^{[2],\star}(\rho),
  \]
  which is exactly \eqref{eq:Delta2-comp}.
  Finally, combining \eqref{eqn:1/16}, \eqref{eqn:2} and \eqref{eq:Delta2-comp}, we have 
  \[
    \Delta'(\rho)\geq\frac{1}{16}(\Delta')^{[2],\star}(\rho)\geq\frac{1}{8}\Delta^{[2],\star}(\rho)>\Delta(\rho).
  \]
  This completes the proof.

\end{proof}
Since \(\mathcal{P}_{g-1}^{+}\) is finite and totally ordered by the colexicographic order \(\prec\), Theorems~\ref{thm:equal-length} and~\ref{thm:different-length} together imply Theorem~\ref{thm:monotonicity-BGW}.

\section*{Use of AI tools}
The mathematical results and proofs in this paper were obtained by the authors. The authors used AI tools to assist with proofreading and editing the manuscript. All suggestions were reviewed by the authors, and all changes to the text were made by the authors themselves. The authors take full responsibility for the content of this paper.

\section*{Conflict of interest and data availability statement}
The authors declare that they have no conflicts of interest. No datasets were generated or analysed during the current study.

\vspace{.2in}
{\em Acknowledgements}.
The authors thank Zhiyuan Wang and Jian Zhou for their collaboration on related work. The second author is grateful to Prof. Jianxun Hu for his encouragement. CY is partially supported by the National Natural Science Foundation of China (No. 12401079) and the Natural Science Foundation of Hubei Province (No. 2026AFA008).

\addcontentsline{toc}{section}{References}
\bibliographystyle{abbrv}
\bibliography{reference}
\end{document}